\documentclass[journal]{IEEEtran}
\usepackage{graphicx}
\usepackage[cmex10]{amsmath}
\usepackage{cases}
\usepackage[tight,footnotesize]{subfigure}
\usepackage{amsthm}
\usepackage{cite}
\usepackage{amssymb}
\usepackage{algorithm}
\usepackage{algorithmic}
\usepackage{multirow}
\usepackage{amsmath}
\usepackage{xcolor}
\usepackage{CJK}
\usepackage{subeqnarray}
\usepackage{cases}
\usepackage{enumerate}
\usepackage{cuted}
\usepackage{booktabs}
\usepackage{mathrsfs}
\usepackage{bbm}
\usepackage{array}
\newcolumntype{M}[1]{>{\centering\arraybackslash}m{#1}}

\newtheorem{theorem}{\hskip\parindent \it Theorem}\newtheorem{remark}{\hskip\parindent\it{Remark}}

\ifCLASSINFOpdf
\else
\fi

\begin{document}
\title{Robust Full-Space Physical Layer Security for STAR-RIS-Aided Wireless Networks: Eavesdropper with Uncertain Location and Channel}
\author{Han Xiao,~\IEEEmembership{Student Member,~IEEE,} Xiaoyan Hu$^*$,~\IEEEmembership{Member,~IEEE,}\\
    Ang Li,~\IEEEmembership{Senior Member,~IEEE,} Wenjie Wang,~\IEEEmembership{Senior Member,~IEEE,} %\\   Zhou Su,~\IEEEmembership{Senior Member,~IEEE,}
    %Feifei Gao,~\IEEEmembership{Fellow,~IEEE,}
    Kun~Yang,~\IEEEmembership{Fellow,~IEEE} %Kai-Kit~Wong,~\IEEEmembership{Fellow,~IEEE},
	
\thanks{This work is supported by the National Natural Science Foundation of China under Grant 62201449, the Young Elite Scientists Sponsorship Program by CAST under Grant No.YESS20230611, the Key R$\&$D Projects of Shaanxi Province under Grant 2023-YBGY-040, and the Qin Chuang Yuan High-Level Innovation and Entrepreneurship Talent Program under Grant QCYRCXM-2022-231. %, and the Si Yuan Scholar Foundation. %The work of Ang Li was supported in part by the Young Elite Scientists Sponsorship Program by CIC under Grant 2021QNRC001 and in part by the Science and Technology Program of Shaanxi Province under Grant 2021KWZ-01.
\emph{(Corresponding author: Xiaoyan Hu.)}}	
\thanks{H. Xiao, X. Hu, A. Li, and Wenjie Wang are with the School of Information and Communications Engineering, Xi'an Jiaotong University, Xi'an 710049, China. (email: hanxiaonuli@stu.xjtu.edu.cn, xiaoyanhu@xjtu.edu.cn,  ang.li.2020@xjtu.edu.cn, wjwang@mail.xjtu.edu.cn).}
%\thanks{F. Gao is with the Institute for Artificial Intelligence and the State Key Lab of Intelligent Technologies and Systems, Beijing National Research Center for Information Science and Technology, Department of Automation, Tsinghua University, Beijing 100084, China (email: feifeigao@ieee.org).}
	%\thanks{K.-K. Wong is with the Department of Electronic and Electrical Engineering, University College London, London WC1E 7JE, U.K. (email: kai-kit.wong@ucl.ac.uk)}
%\thanks{Z. Su is with the School of Cyber Science and Engineering, Xi'an Jiaotong University, Xi'an 710049, China. (email: zhousu@xjtu.edu.cn).}
\thanks{K. Yang is with the School of Computer Science and Electronic Engineering, University of Essex, Colchester CO4 3SQ, U.K. (e-mail: kunyang@essex.ac.uk).}
}
\maketitle

\begin{abstract}
A robust full-space physical layer security (PLS) transmission scheme is proposed in this paper considering the full-space wiretapping challenge of wireless networks supported by simultaneous transmitting and reflecting reconfigurable intelligent surface (STAR-RIS). Different from the existing schemes, the proposed PLS scheme takes account of the uncertainty on the eavesdropper's position within the 360$^\circ$ service area offered by the STAR-RIS. Specifically, the large system analytical method is utilized to derive the asymptotic expression of the average security rate achieved by the security user, considering that the base station (BS) only has the statistical information of the eavesdropper's channel state information (CSI) and the uncertainty of its location. To evaluate the effectiveness of the proposed PLS scheme, we first formulate an optimization problem aimed at maximizing the weighted sum rate of the security user and the public user. This optimization is conducted under the power allocation constraint,  and some practical limitations for STAR-RIS implementation, through jointly designing the active and passive beamforming variables. A novel iterative algorithm based on the minimum mean-square error (MMSE) and cross-entropy optimization (CEO) methods is proposed to effectively address the established non-convex optimization problem with discrete variables. Simulation results indicate that the proposed robust PLS scheme can effectively mitigate the information leakage across the entire coverage area of the STAR-RIS-assisted system, leading to superior performance gain when compared to benchmark schemes encompassing traditional RIS-aided scheme. %These validate the efficacy of the proposed robust PLS scheme and the proposed iterative algorithm.
\end{abstract}
\begin{IEEEkeywords}
	Robust physical layer security, simultaneous transmitting and reflecting RIS (STAR-RIS), uncertain location, cross-entropy optimization (CEO).
\end{IEEEkeywords}
\IEEEpeerreviewmaketitle

\vspace{-1mm}
\section{Introduction}\label{sec:S1}
The ongoing advancements in wireless communication technologies have led to a growing demand for enhanced wireless security measures in both the existing 5G networks and the upcoming 6G networks. The physical layer security (PLS) based on the information theory framework has emerged as a promising solution, which harnesses the inherent unpredictability of physical medium, especially the random variations of the legitimate and potential wiretap channels, to protect the security of wireless information transmissions \cite{liang2009information, shiu2011physical, zhiselfpowered2024}.
The massive multiple-input multiple-output (MIMO) technique, as a pivotal technology for 5G and promising 6G, offers an effective way to implement high-performance PLS \cite{larsson2014massive, kapetanovic2015physical, zhu2017analysis, xiao2018opportunistic}. Due to the fact that the propagation channel vectors belonging to different users gradually become orthogonal as the number of antennas increases indefinitely, thereby the technology of massive MIMO can help mitigate the impact of multi-user interference. However, the continuous increase in the size of antenna arrays poses numerous challenges for traditional phased array-based MIMO systems, including elevated costs, heightened hardware complexity, substantial insertion losses, as well as unacceptable energy consumption. These obstacles will definitely impede the widespread applications of the massive MIMO systems in 5G and 6G networks \cite{shlezinger2021dynamic}.
\subsection{Related Works}
Reconfigurable intelligent surfaces (RISs) have recently gained prominence as a crucial technology in the context of 6G communications, providing a cost-efficient and hardware-efficient strategy to address the challenges faced by massive MIMO systems. By modifying the resonant structure of its passive metasurface components, RIS has the capability to control the electromagnetic properties of incoming signals, thereby reshaping the wireless propagation environment and improving the quality of communication channels \cite{huang2019reconfigurable, pan2021reconfigurable, sunmulti2024}.
Hence, the potentials of the RIS have attracted great attention, which has been applied and verified in various communication systems, such as cell-free networks \cite{macooperative2023}, satellite-terrestrial communications \cite{TAES2022_ZLin_RIS}, Internet of Things (IoT) systems \cite{zhangIRS2024}, simultaneous
wireless information and power transfer (SWIPT)\cite{anexploiting2024}, PLS \cite{cui2019secure, hong2020artificial, xu2021intelligent,  wang2020energy}. %tang2021securing,
Specifically, RIS is first utilized to improve the legitimate user' PLS performance in \cite{cui2019secure}. To assess the efficacy of the RIS in bolstering PLS performance, a challenging scenario is examined, where the eavesdropping channel is closely aligning with the legitimate direct channel and the eavesdropper possesses a superior channel link. The RIS facilitates the adjustment of phase-shifts for the incoming signals according to precise channel state information (CSI) of the eavesdropper, enabling the non-coherent summation of signals reflected by the RIS and those transmitted directly to the eavesdropper.
This process can effectively weaken the signal strength received by the eavesdropper. However, it is  difficult to achieve the perfect CSI of the eavesdropper in practical scenarios. In addition, eavesdroppers may be much more powerful than we imagine. For example, eavesdroppers exhibit superior hardware resources and computational capacities compared to authorized users, exemplified by their possession of antenna elements and adept interference mitigation capabilities. Hence, it is necessary to investigate other advanced technologies to deal with the cases with powerful eavesdroppers.

To further enhance the PLS performance and provide robust PLS schemes in scenarios with uncertain information of eavesdroppers, the artificial interference has been considered as an effective way. A set of interference sources are introduced in the RIS-aided secure communication systems, with the objective of degrading the reception quality of eavesdroppers, especially for those with higher capabilities. In particular, the authors of \cite{hong2020artificial} employ the artificial noise (AN) emitted from the base station (BS) under the assistance of the RIS, to counteract eavesdropping by multi-antenna eavesdropper on the communications between the BS and legitimate users. The simulation results have shown the significant potentials of the combination of the RIS and AN in enhancing the security performance. In \cite{xu2021intelligent}, RIS operates as the source of interference to produce jamming signals, causing a degradation in the quality of the received signals at the eavesdroppers. Essentially, the incident signal is remodulated into random jamming signals by the RIS, to interfere potential eavesdroppers. Additionally, by taking advantage of the full-duplex technology, mobile users can transmit jamming signals while receiving, which can also help to disrupt the eavesdropping attempts \cite{wang2020energy}.%tang2021securing, 

It is worth noting that the traditional RIS (\textcolor{blue}{including passive RIS and active RIS \cite{zhang2023beamforming})} can only implement the reflected modulation for the incident signals in the existing RIS-aided PLS systems, which means the transceiver terminal equipment must be located on the same side as the RIS. This will significantly limit the coverage area of the wireless networks and the flexibility in deploying the RIS. An more advanced RIS known as simultaneously transmitting and reflecting RIS (STAR-RIS) has been introduced in \cite{liu2021star} to address the aforementioned obstacles, due to the fact that STAR-RIS can divide the incoming signal into two components, with one portion being reflected back on the same side as the incident signal and the other portion being transmitted to the opposite side. Consequently, STAR-RIS can create a comprehensive smart propagation environment covering $360^\circ$ area, comprising a reflection region (R region) and a transmission region (T region). In the R region, communication services are facilitated through reflection modulation capability of the STAR-RIS, while the transmission modulation capacity is employed to enhance services in the T region.

Hence, STAR-RIS demonstrates great potentials in wireless networks.  Note that several PLS transmitting schemes in STAR-RIS-assisted communication systems have been investigated \cite{li2022enhancing, sheng2023performance, han2022artificial, wen2024star, zhou2023robust, shen2024star}. %li2023reliability,, zhang2023star
Specifically, \cite{li2022enhancing, sheng2023performance} centre on assessing the security efficacy of STAR-RIS-enhanced non-orthogonal multiple access (NOMA) networks and cognitive radio NOMA wireless networks. %li2023reliability, 
The authors of \cite{han2022artificial} initially utilize the AN to assist the PLS in the STAR-RIS-aided wireless networks and fully explores the interrelationship between the power allocation of the AN and the number of RIS elements as well as the number of antennas. The researchers in \cite{wen2024star} introduce a novel PLS scheme integrated with full-duplex technique in the  STAR-RIS-enhanced communication systems. In this setup, a subset of antennas at Bob is designated for signal reception from Alice, while the remaining antennas are designated for transmitting jamming signals to disrupt potential eavesdroppers within the transmission region, leveraging the transmission capabilities of the STAR-RIS. It is important to highlight that the STAR-RIS's reflective modulation is efficiently employed to eliminate self-interference within the full-duplex receiver. Furthermore, \cite{zhou2023robust} and \cite{shen2024star} respectively present the high-robustness STAR-RIS-aided PLS transmission scheme considering the practical scenario where BSs face challenges in obtaining accurate CSI of potential eavesdroppers.

\vspace{-2mm}
\subsection{Motivations and Contributions}
Actually, the above  PLS schemes in STAR-RIS-aided wireless networks generally have the following deficiencies: (\romannumeral 1) These PLS schemes usually assume that the eavesdropper's activity region (R region or T region) within the service area of the STAR-RIS  has been pre-determined, which enables the systems to effectively address the eavesdropping risks by designing appropriate security measures. It is worth noting that this assumption is impractical due to the fact that smart eavesdroppers in practice can  fully exploit the full-space coverage feature of the STAR-RIS to augment their capacity for intercepting confidential information. Specifically, they can freely  move  within R region and T region to enhance the probability of successful wiretapping, posing a tough challenge for current PLS schemes in maintaining information security.
(\romannumeral 2) These existing PLS schemes can only ensure the half-space wireless communication security, since they focus on guaranteeing security within the R region or T region considering a fixed eavesdropper located in the specific region. Note that the STAR-RIS's full-space coverage makes the communication systems assisted by STAR-RIS encounter more severe inherent security issues than conventional RIS-aided systems, since eavesdroppers can conduct $360^\circ$ monitoring for the communications between the BS and legitimate users.

In response to the shortcomings of the existing PLS solutions mentioned above, taking full account of the pervasive eavesdropping risks faced by the STAR-RIS-aided wireless networks, a full-space PLS scheme is proposed in \cite{zhangsecurity2023}, where two eavesdroppers are assumed to respectively situate in R region and T region and carry out wiretapping on users within these areas. Although this PLS transmission scheme can obtain the unique full-space communication security of the STAR-RIS-aided communications, it is designed assuming that the BS possesses the precise  CSI of eavesdroppers. This assumption is unrealistic, as intelligent eavesdroppers will make every effort to conceal themselves to increase their chances of successful interception, making it challenging for the BS to acquire the perfect CSI. Hence, the main motivations for conducting this study arise from the urgent requirement to address the challenges mentioned above. To this end, a robust full-space PLS transmitting scheme is proposed to protect the communication security of the STAR-RIS-supported wireless networks.  Our primary contributions are outlined as follows:
\begin{itemize}
\vspace{1mm}
\item\textbf{\textit{Robust STAR-RIS-assisted PLS Scheme Considering the Uncertain Position of Eavesdropper:}}
A novel robust full-space PLS transmission scheme is proposed for the STAR-RIS-aided communication systems. In contrast to existing PLS schemes, the proposed PLS scheme initially takes into account the uncertainty on eavesdropper's location within the full-space coverage area of the STAR-RIS-aided networks. This scheme can effectively mitigate information leakage for legitimate users from both R and T regions by carefully optimizing the passive reflection and transmission coefficients of the STAR-RIS, along with the active beamforming employed at the BS.
\vspace{1mm}
\item\textbf{\textit{Problem Formulation under Practical Constraints with Closed-form Derivations of PLS Performance Indicators:}}
The large system analytical technique is leveraged to derive the closed-form expressions of the average security rate, assuming that the BS has access solely to the statistical CSI pertaining to the eavesdropper, along with the uncertainty of eavesdropper's location in either R region or T region. Then, the optimization problem under discrete phase-shifts and transmitting power constraints is formulated to maximize the weighted sum of the average security rate at the security user and the achievable rate of the public user. This is achieved through joint design of active and passive beamforming variables.
\vspace{1mm}	
\item\textbf{\textit{Effective Alternating Algorithm with Guaranteed Convergence and Low Computational Complexity:}}
Note that the formulated problem is shown to be a non-convex optimization problem considering the high coupling among variables and the non-convexity of constraints and the objective function. To handle this problem, the alternative strategy is leveraged to divide the optimization problem into the active beamforming subproblem and the passive beamforming subproblem. For the active beamforming subproblem, the minimum mean-square error (MMSE) and subgradient method are utilized to tackle this optimization problem with semi-closed solutions. The passive beamforming problem, involving discrete phase-shift constraints and equal amplitude constraints, is efficiently addressed by employing the cross-entropy optimization (CEO) method, which offers a closed-form solution. The designed algorithm's convergence can be ensured and verified by the simulation results showcasing convergence curves. Additionally, the computational complexity of the iterative algorithm is calculated, which is notably less than that of many existing optimization algorithms.
\vspace{1mm}
\item \textbf{\textit{Substantial Performance Improvement:}}
Extensive simulations are implemented and the obtained results indicate that the proposed robust full-space PLS scheme can bring considerable performance gain by
 comparing with baseline schemes including two benchmark algorithms (zero-forcing algorithm and semi-definite relaxation algorithm) as well as the traditional RIS-aided scheme. Additionally, the presented simulation results also demonstrate that the proposed robust full-space PLS scheme can effectively suppresses the information leakage throughout the entire coverage region of the STAR-RIS-assisted wireless networks. These results clearly validate the efficacy of the proposed PLS scheme and iterative algorithm.
\end{itemize}

The subsequent sections of this paper are structured as below. The details of system model, channel model and signal model are shown in Section \ref{sec:S2}. Section \ref{sec:S3} derives the asymptotic expression of the average security rate utilizing the large system analytic technology and formulates the optimization problem. The designed iterative algorithm is given in Section \ref{sec:S4}, including complexity and convergence analysis. Extensive simulations are performed to verify the effectiveness of the proposed robust full-space PLS scheme and the proposed iterative algorithm in Section \ref{sec:S5}. Section \ref{sec:S6} conduct the conclusion for the entire study.

\textit{Notation:}  $(\cdot)^T$ and $(\cdot)^H$ represent transpose and conjugate transpose, respectively. $\operatorname{Diag}(\mathbf{a})$ is a diagonal matrix where the diagonal elements are specified by the vector $\mathbf{a}$. $|\cdot|$ and $\|\cdot\|$ indicate the modulus value of a complex scalar and  the Euclidean norm of a complex vector, respectively. $\mathbb{C}^{M\times1}$ denotes the set of $M\times1$ complex vectors. $x\sim \mathcal{CN}(a, b)$ represents a circularly symmetric complex Gaussian random variable denoted by $x$, characterized by a mean of $a$ and a variance of $b$. Operator $\leftarrow$ denotes the assignment operation. Operator $[x]^{+}$ represents the operation of $\max(0, x)$. $\mathbf{I}_{N\times N}$ denotes the $N\times N$ identity matrix.
\section{System model}\label{sec:S2}
\begin{figure}[ht]
	\centering
	\includegraphics[scale=0.38]{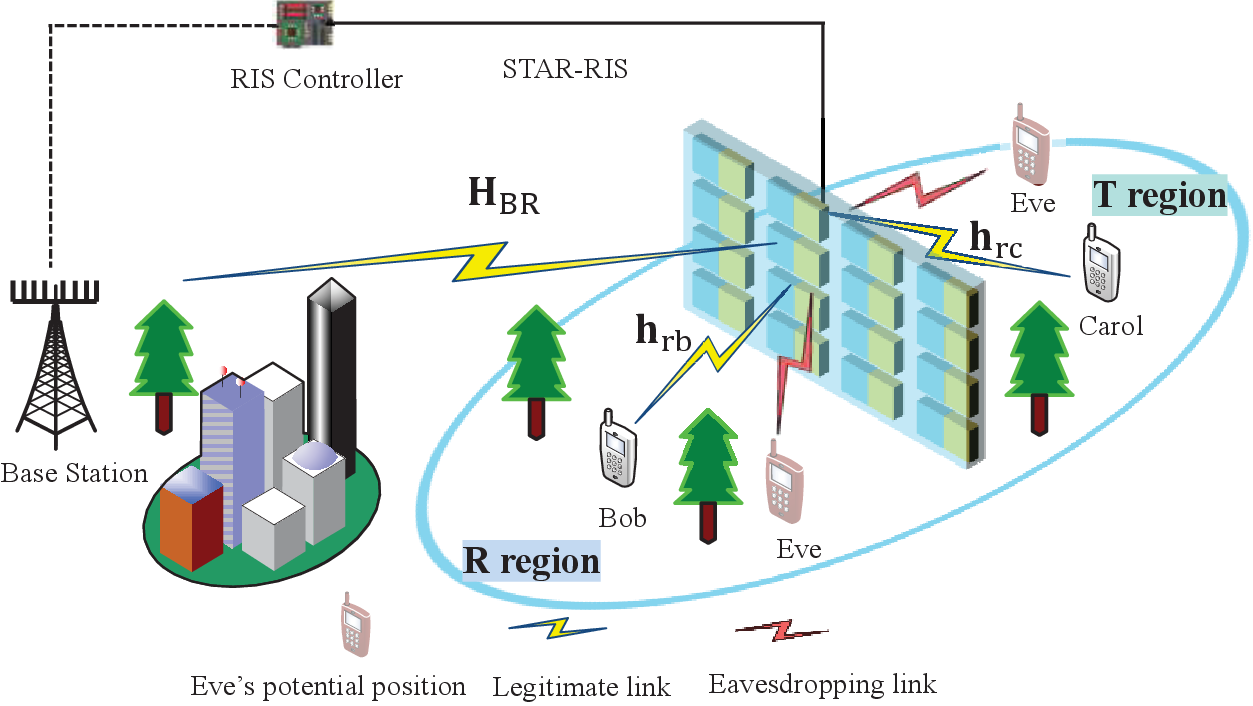}\\
	\caption{Modeling System for PLS Architecture Assisted by STAR-RIS.}\label{fig:scenario}
\end{figure}
The system model of the STAR-RIS-assisted PLS architecture is illustrated in Fig. \ref{fig:scenario}, comprising a BS with $N_\mathrm{t}$ antennas, a STAR-RIS with $M$ elements, and three distinct users: the security user Bob, the public user Carol, and the eavesdropping user Eve. It is assumed that all users utilize a single-antenna functioning in half-duplex mode. The STAR-RIS is positioned in close proximity to users so as to improve the end-and-end communication qualities between the BS and users. Note that the service area will be divided into two regions by the STAR-RIS: the R region and the T region. Users in the R region will benefit from communication services facilitated by the reflection signals of the STAR-RIS, whereas users in the T region will access communication services through the transmission signals of the STAR-RIS. Without loss of generality, it is assumed that Bob, the security user, is situated within the R region, whereas Carol, the public user, utilizes the communication service within the T region.
The energy splitting (ES) protocol \cite{mu2021simultaneously} is considered to implement the STAR-RIS in this paper, indicating that each element of STAR-RIS can reflect and transmit the signals simultaneously. In addition, physical obstructions like buildings and trees hinder the direct links between the BS and the users.

To improve the  capacity for eavesdropping the communications between the BS and Bob,  intelligent Eve endeavours to conceal its presence and moves freely within the full-space served by the STAR-RIS-aided wireless networks. Thus, it is difficult for BS to acquire the accurate region and the perfect CSI of Eve, which makes it challenging to further design and perform the corresponding beamforming strategy to decrease the signal quality at Eve. Next, we will present the channel model and signal model.

\subsection{Channel model}
The connections between the STAR-RIS and the BS ($\mathbf{H}_\mathrm{BR}$) as well  between the STAR-RIS and users ($\mathbf{h}_{\mathrm{r}\zeta}$) are respectively modeled as
%Rayleigh fading channel. Specifically, the channels can be respectively given by
\begin{align}
	&\mathbf{H}_\mathrm{BR}=\sqrt{l_\mathrm{BR}}\mathbf{G}_\mathrm{BR}\in\mathbb{C}^{M\times N_t},\\
	&\mathbf{h}_{\mathrm{r}\zeta}=\sqrt{l_{\mathrm{r}\zeta}}\mathbf{g}_{\mathrm{r}\zeta}\in\mathbb{C}^{M\times 1}, ~\zeta\in\{\mathrm{b, c, e}\},
\end{align}
where $l_\mathrm{BR}=\frac{\rho}{d_\mathrm{BR}^{\alpha}}$ and $l_{\mathrm{r}\zeta}=\frac{\rho}{d_{\mathrm{r}\zeta}^{\alpha}}$, with $\rho$ denoting the reference power gain at a distance of $1$ m, $\alpha$ representing the path-loss exponent. $d_\mathrm{BR}$ and $d_{\mathrm{r}\zeta}$ represent the spatial separation between the BS and the STAR-RIS, and between the STAR-RIS and the users, respectively. %and $d_\mathrm{BR}$ is the spatial distance between the BS and the STAR-RIS.
The matrices $\mathbf{G}_\mathrm{BR}\in\mathbb{C}^{M\times N_t}$ and $\mathbf{g}_{\mathrm{r}\zeta}\in\mathbb{C}^{M\times 1}$ represent the small-scale Rayleigh fading coefficients. Each entry of these matrices follows an independent identically distributed (i.i.d.) pattern, conforming to a complex Gaussian distribution with a mean of zero and variance of one.
\subsection{Signal model}
On the basis of the analysis above, the received signal at Bob and Carol can be expressed as
\begin{align}
	&y_\mathrm{b}=\mathbf{h}_\mathrm{rb}^H\mathbf{U}_\mathrm{r}\mathbf{H}_\mathrm{BR}\left(\mathbf{w}_\mathrm{b}s_\mathrm{b}+\mathbf{w}_\mathrm{c}s_\mathrm{c}\right)+n_\mathrm{b},\\
	&y_\mathrm{c}=\mathbf{h}_\mathrm{rc}^H\mathbf{U}_\mathrm{t}\mathbf{H}_\mathrm{BR}\left(\mathbf{w}_\mathrm{b}s_\mathrm{b}+\mathbf{w}_\mathrm{c}s_\mathrm{c}\right)+n_\mathrm{c},
\end{align}
where $s_\mathrm{b}$ and $s_\mathrm{c}$ denote the transmitted signal from the BS to Bob and Carol, respectively. The precoding vectors allocated by the BS for Bob and Carol are denoted as $\mathbf{w}_\mathrm{b} \in\mathbb{C}^{N_t\times 1}$ and $\mathbf{w}_\mathrm{c} \in\mathbb{C}^{N_t\times 1}$, respectively. $n_\mathrm{b} \sim \mathcal{CN}(0, \sigma_\mathrm{b}^2)$ and $n_\mathrm{c} \sim \mathcal{CN}(0, \sigma_\mathrm{c}^2)$ are the additive white Gaussian noise (AWGN) with $\sigma_\mathrm{b}^2$ and $\sigma_\mathrm{c}^2$ being the noise power. $\mathbf{U}_\mathrm{r}$ and $\mathbf{U}_\mathrm{t}$ represent the reflection and transmission coefficients of the STAR-RIS, which are given by
\begin{itemize}
		\vspace{2mm}
	\item $\mathbf{U}_\mathrm{r}=\operatorname{Diag}\left\{\sqrt{\beta_1^{\mathrm{r}}}e^{j\phi^{\mathrm{r}}_1}, \cdots, \sqrt{\beta_M^{\mathrm{r}}}e^{j\phi^{\mathrm{r}}_M}\right\},$
	\vspace{2mm}
	\item $\mathbf{U}_\mathrm{t}=\operatorname{Diag}\left\{\sqrt{\beta_1^{\mathrm{t}}}e^{j\phi^{\mathrm{t}}_1}, \cdots, \sqrt{\beta_M^{\mathrm{t}}}e^{j\phi^{\mathrm{t}}_M}\right\},$	
		\vspace{2mm}
\end{itemize}
with $\beta_m^{\mathrm{r}}+\beta_m^{\mathrm{t}}=1, ~\textcolor{blue}{\phi^{\mathrm{r}}_m,~ \phi^{\mathrm{t}}_m \in(0, 2\pi]},~ \forall m\in\mathcal{M}\triangleq\{1, \cdots, M\}$. Note that $\beta_m^{\mathrm{r}}, \beta_m^{\mathrm{t}}\in (0,1)$,
%for the MS protocol, $\beta_m^{\mathrm{r, MS}}, \beta_m^{\mathrm{t, MS}}\in \{0, 1\}$.

When Eve traverses in the service region of the STAR-RIS-aided systems, it may perform the wiretap for the signals in different region. In other words, Eve may engage in eavesdropping activities within the R region or T region.
In order to represent Eve's position region, a Bernoulli variable $b$ is defined. Specifically, when $b=1$ with probability $P_1$, it signifies that Eve is positioned in the R region of the STAR-RIS. Conversely, when $b=0$ with probability $P_0=1-P_1$, it indicates that Eve is positioned in the T region of the STAR-RIS. Hence, the signals received by the Eve with non-fixed location can be expressed as
\begin{align}
y_\mathrm{e}=\mathbf{h}_\mathrm{re}^H(b\mathbf{U}_\mathrm{r}+(1-b)\mathbf{U}_\mathrm{t})\mathbf{H}_\mathrm{BR}(\mathbf{w}_\mathrm{b}s_\mathrm{b}+\mathbf{w}_\mathrm{c}s_\mathrm{c})+n_\mathrm{e},
\end{align}
where $n_\mathrm{e} \sim \mathcal{CN}(0, \sigma_\mathrm{e}^2)$ denotes the AWGN at Eve with $\sigma_\mathrm{e}^2$ being the noise power.

\section{Analysis on PLS and Problem Formulation}\label{sec:S3}
\subsection{PLS Analysis}
According to the analysis in Section \ref{sec:S2}, the achievable rate at Bob and Carol can respectively given by
\begin{align}
&	R_\mathrm{b}=\log_2\left(1+\frac{\left|\mathbf{h}_\mathrm{rb}^H\mathbf{U}_\mathrm{r}\mathbf{H}_\mathrm{BR}\mathbf{w}_\mathrm{b}\right|^2}{\left|\mathbf{h}_\mathrm{rb}^H\mathbf{U}_\mathrm{r}\mathbf{H}_\mathrm{BR}\mathbf{w}_\mathrm{c}\right|^2+\sigma_\mathrm{b}^2}\right),\\
  &  R_\mathrm{c}=\log_2\left(1+\frac{\left|\mathbf{h}_\mathrm{rc}^H\mathbf{U}_\mathrm{t}\mathbf{H}_\mathrm{BR}\mathbf{w}_\mathrm{c}\right|^2}{\left|\mathbf{h}_\mathrm{rc}^H\mathbf{U}_\mathrm{t}\mathbf{H}_\mathrm{BR}\mathbf{w}_\mathrm{b}\right|^2+\sigma_\mathrm{c}^2}\right).
\end{align}
The eavesdropping rate of Eve to communications between the BS and Bob can be expressed as
\begin{align}
&R_\mathrm{eb}=\notag\\	&\log_2\left(1+\frac{\left|\mathbf{h}_\mathrm{re}^H(b\mathbf{U}_\mathrm{r}+(1-b)\mathbf{U}_\mathrm{t})\mathbf{H}_\mathrm{BR}\mathbf{w}_\mathrm{b}\right|^2}{\left|\mathbf{h}_\mathrm{re}^H(b\mathbf{U}_\mathrm{r}+(1-b)\mathbf{U}_\mathrm{t})\mathbf{H}_\mathrm{BR}\mathbf{w}_\mathrm{c}\right|^2+\sigma^2_\mathrm{e}}\right).
\end{align}
Therefore, the security rate for Bob is given by
\begin{align}
	R_\mathrm{b}^\mathrm{s}=[R_\mathrm{b}-R_\mathrm{eb}]^{+}.
\end{align}

Recall that it is challenging for the BS to obtain accurate knowledge of Eve's location region and perfect CSI, given Eve's efforts to remain hidden and move freely throughout the entire coverage area of the STAR-RIS-assisted wireless network. The average security rate over $b$ and $\mathbf{h}_\mathrm{re}$ is selected to denote Bob's secure performance in STAR-RIS-aided communication system, which is given by
 \begin{align}
&\tilde{R}_\mathrm{b}^\mathrm{s}=\mathbb{E}_{\{b, \mathbf{h}_\mathrm{re} \}}(R_\mathrm{b}^\mathrm{s})=R_\mathrm{
b}-\mathbb{E}_{\{b, \mathbf{h}_\mathrm{re}\}}(R_\mathrm{eb}),
\end{align}
where
\begin{align}
&\mathbb{E}_{\{b, \mathbf{h}_\mathrm{re}\}}(R_\mathrm{eb})\notag\\
&=P_1\mathbb{E}_{\{\mathbf{h}_\mathrm{re}\}}\Bigg(\log_2\bigg(1+\frac{\left|\mathbf{h}_\mathrm{re}^H\mathbf{U}_\mathrm{r}\mathbf{H}_\mathrm{BR}\mathbf{w}_\mathrm{b}\right|^2}{\left|\mathbf{h}_\mathrm{re}^H\mathbf{U}_\mathrm{r}\mathbf{H}_\mathrm{BR}\mathbf{w}_\mathrm{c}\right|^2+\sigma^2_\mathrm{e}}\bigg)\Bigg)+\notag\\
&~~~P_0\mathbb{E}_{\{\mathbf{h}_\mathrm{re}\}}\Bigg(\log_2\bigg(1+\frac{\left|\mathbf{h}_\mathrm{re}^H\mathbf{U}_\mathrm{t}\mathbf{H}_\mathrm{BR}\mathbf{w}_\mathrm{b}\right|^2}{\left|\mathbf{h}_\mathrm{re}^H\mathbf{U}_\mathrm{t}\mathbf{H}_\mathrm{BR}\mathbf{w}_\mathrm{c}\right|^2+\sigma^2_\mathrm{e}}\bigg)\Bigg).
\end{align}
Actually, the precise mathematical formulation for the $\mathbb{E}_{\{b, \mathbf{h}_\mathrm{re}\}}(R_\mathrm{eb})$ is deemed intractable due to the incorporation of the random variable $\mathbf{h}_\mathrm{re}$ within the eavesdropping rate expression. This feature adds complexity to ascertaining the probability distribution of the rate expression involving the random variable $\mathbf{h}_\mathrm{re}$. To address this issue, the large system analytic technique in \cite{evans00} is adopted, which is a commonly employed method for evaluating the performance limitations of wireless communication systems \cite{wang2021intelligent, Xiao2024simultaneously, Xiao2024STAR}. Specifically, it is presumed that the STAR-RIS comes with a substantial amount of passive elements, therefore, we have
\begin{align}
&\lim_{M\rightarrow\infty}\frac{\mathbf{h}_\mathrm{re}^H\mathbf{U}_\mathrm{r}\mathbf{H}_\mathrm{BR}\mathbf{w}_\mathrm{b}\mathbf{w}_\mathrm{b}^H\mathbf{H}_\mathrm{BR}^H\mathbf{U}_\mathrm{r}^H\mathbf{h}_\mathrm{re}}{M}\notag\\
&\stackrel{(a)}{\rightarrow}\frac{l_\mathrm{re}\operatorname{Tr}(\mathbf{U}_\mathrm{r}\mathbf{H}_\mathrm{BR}\mathbf{w}_\mathrm{b}\mathbf{w}_\mathrm{b}^H\mathbf{H}_\mathrm{BR}^H\mathbf{U}_\mathrm{r}^H)}{M}\notag\\
&=\frac{l_\mathrm{re}\left\|\mathbf{U}_\mathrm{r}\mathbf{H}_\mathrm{BR}\mathbf{w}_\mathrm{b}\right\|^2}{M},
\end{align}
where $(a) $ is obtained because of the corollary in \cite[Corollary 1]{evans00}. Note that, by conducting such procedures, the randomness introduced by $\mathbf{h}_\mathrm{re}$ can be eliminated in $\mathbb{E}_{\{b, \mathbf{h}_\mathrm{re}\}}(R_\mathrm{eb})$. Hence, the asymptotic expression for the average security rate can be achieved \textcolor{blue}{\footnote{\textcolor{blue}{ Note that the derived PLS model is capable of effectively ensuring information security even in the presence of multiple unauthorized eavesdroppers. This is because the secure rate model is derived using the statistical eavesdropper's CSI, which means that all potential eavesdropping channels are taken into account within the secrecy rate model. Consequently, even in extreme cases involving multiple eavesdroppers, the PLS scheme can still provide effective protection. %The simulation results in Fig. \ref{fig:R_eb_vs_channel} further verify this conclusion.
Actually, the proposed PLS model cannot fully characterize the security performance in the multiple eavesdropper scenarios considering that the potential collaborative behaviors, and the complexity introduced by different channel conditions. However, the derivation techniques presented in this paper can be extended to develop PLS models for multi-eavesdropper scenarios, providing a foundation for further investigations.}}}, which is derived as
\begin{align}
\tilde{R}_\mathrm{b}^\mathrm{sa}=&R_\mathrm{b}-P_1\log_2\left(1+\frac{l_\mathrm{re}\left\|\mathbf{U}_\mathrm{r}\mathbf{H}_\mathrm{BR}\mathbf{w}_\mathrm{b}\right\|^2}{l_\mathrm{re}\left\|\mathbf{U}_\mathrm{r}\mathbf{H}_\mathrm{BR}\mathbf{w}_\mathrm{c}\right\|^2+\sigma^2_\mathrm{e}}\right)\notag\\
&-P_0\log_2\left(1+\frac{l_\mathrm{re}\left\|\mathbf{U}_\mathrm{t}\mathbf{H}_\mathrm{BR}\mathbf{w}_\mathrm{b}\right\|^2}{l_\mathrm{re}\left\|\mathbf{U}_\mathrm{t}\mathbf{H}_\mathrm{BR}\mathbf{w}_\mathrm{c}\right\|^2+\sigma^2_\mathrm{e}}\right).
\end{align}
Therefore, the asymptotic result of $\tilde{R}_\mathrm{b}^\mathrm{s}$, i.e., $\tilde{R}_\mathrm{b}^\mathrm{sa}$, will be utilized to represent the security rate for variable optimization and algorithm design in next sections.
\vspace{-2mm}\subsection{Problem Formulation}
The optimization problem is established in this section. In particular, the aim of this paper is to optimize the weighted sum of the average security rate at Bob and the achievable rate at Carol through the joint design of the active and passive beamforming variables. It is worth noting that the STAR-RIS utilizes the discrete phase-shifts to modulate the incident signals in this paper, taking into account the hardware constraints in real-world scenarios. In this case, the formulated optimization problem can be expressed as
\begin{subequations}\label{eq_oringi_opti}
	\begin{align}
		&	\max_{\mathbf{w}_\mathrm{b}, \mathbf{w}_\mathrm{c},\mathbf{U}_\mathrm{r},\mathbf{U}_\mathrm{t}}\omega_1\tilde{R}_\mathrm{b}^\mathrm{sa}+\omega_2R_\mathrm{c}\notag\\
		&\qquad	\text{s.t.}~\mathbf{w}_\mathrm{b}^H\mathbf{w}_\mathrm{b}+\mathbf{w}_\mathrm{c}^H\mathbf{w}_\mathrm{c}\leq P_\mathrm{tmax},\label{eq_oringi_opti_1}\\
		&\qquad\quad \beta_{m}^{\mathrm{r}}, \beta_{m}^{\mathrm{t}}\in\mathcal{B}, ~\forall m\in\mathcal{M},\label{eq_oringi_opti_2} \\
		&\qquad\quad\beta_{m}^{\mathrm{r}}+\beta_{m}^{\mathrm{t}}=1, \forall m\in\mathcal{M}, \mathrm{MS}\},\label{eq_oringi_opti_3} \\
		&\qquad\quad \phi_m^{\mathrm{r}}, \phi_m^{\mathrm{r}}\in\{\frac{2\pi}{2^{\lambda}},2\frac{2\pi}{2^{\lambda} }, \cdots, 2^{\lambda}\frac{2\pi}{2^{\lambda}}\},\label{eq_oringi_opti_4}
	\end{align}
\end{subequations}
where $\omega_1$, $\omega_2\in(0,1]$ denote the weighted factors; $P_\mathrm{tmax}$ represents the maximum transmitting power at the BS; \textcolor{blue}{Given that the ES protocol is utilized to implement the STAR-RIS,} $\mathcal{B}\triangleq(0,1]$; $\lambda$ denotes the number of quantization bits. Actually, it is difficult to address this problem because of the non-convex objective function and equal constraints for amplitudes in \eqref{eq_oringi_opti_3}, the existing discrete variables, as well as the significant coupling among variables. To handle this problem, the alternative strategy is first leveraged to divided the original optimization problem into two subproblems, i.e., active beamforming subproblem and passive beamforming subproblem. Then, a novel iterative algorithm based on the MMSE and the CEO approaches is proposed to effectively address the formulated problem.
\section{Algorithm Design}\label{sec:S4}
\subsection{Algorithm Design}
\subsubsection{Active beamforming design} This section focus on optimizing the active beamforming variables, i.e., $\mathbf{w}_\mathrm{b}$ and $\mathbf{w}_\mathrm{c}$, with the given $\mathbf{U}_\mathrm{r}$ and $\mathbf{U}_\mathrm{t}$. Specifically, the subproblem associated with active beamforming variables is given by
\begin{subequations}\label{eq_active}
	\begin{align}
		&	\max_{\mathbf{w}_\mathrm{b}, \mathbf{w}_\mathrm{c}}~\omega_1\tilde{R}_\mathrm{b}^\mathrm{sa}+\omega_2R_\mathrm{c}\notag\\
		&\quad	\text{s.t.}~\mathbf{w}_\mathrm{b}^H\mathbf{w}_\mathrm{b}+\mathbf{w}_\mathrm{c}^H\mathbf{w}_\mathrm{c}\leq P_\mathrm{tmax}. \label{eq_active_1}
	\end{align}
\end{subequations}
Note that the subproblem \eqref{eq_active} is a non-convex optimization problem, since the objective function is a non-convex function with respect to (w.r.t) $\mathbf{w}_\mathrm{b}$ and $\mathbf{w}_\mathrm{c}$. In order to solve this problem, we first transform $\tilde{R}_\mathrm{b}^\mathrm{sa}$ as the following form:
\begin{align}\label{eq_Rb_trans}
&\hspace{-2mm}\tilde{R}_\mathrm{b}^\mathrm{sa}	=\frac{1}{\ln2}\underbrace{\ln\Big(1+\frac{\left|\mathbf{h}_\mathrm{rb}^H\mathbf{U}_\mathrm{r}\mathbf{H}_\mathrm{BR}\mathbf{w}_\mathrm{b}\right|^2}{\left|\mathbf{h}_\mathrm{rb}^H\mathbf{U}_\mathrm{r}\mathbf{H}_\mathrm{BR}\mathbf{w}_\mathrm{c}\right|^2+\sigma^2_\mathrm{b}}\Big)}_{f_1}+\frac{P_1}{\ln2}\times\notag\\
&\hspace{-2mm}\underbrace{\ln\Big(1+\frac{l_\mathrm{re}\left\|\mathbf{U}_\mathrm{r}\mathbf{H}_\mathrm{BR}\mathbf{w}_\mathrm{c}\right\|^2}{\sigma^2_\mathrm{e}}\Big)}_{f_2}+\underbrace{\ln\Big(1+\frac{l_\mathrm{re}\left\|\mathbf{U}_\mathrm{t}\mathbf{H}_\mathrm{BR}\mathbf{w}_\mathrm{c}\right\|^2}{\sigma^2_\mathrm{e}}\Big)}_{f_3}\times\notag\\
	&\hspace{-2mm}\frac{P_0}{\ln2}\underbrace{-\ln\Big(1+\frac{l_\mathrm{re}(\left\|\mathbf{U}_\mathrm{r}\mathbf{H}_\mathrm{BR}\mathbf{w}_\mathrm{c}\right\|^2+\left\|\mathbf{U}_\mathrm{r}\mathbf{H}_\mathrm{BR}\mathbf{w}_\mathrm{b}\right\|^2)}{\sigma^2_\mathrm{e}}\Big)}_{f_4}\frac{P_1}{\ln2}\notag\\
	&\hspace{-2mm}\underbrace{-\ln\Big(1+\frac{l_\mathrm{re}(\left\|\mathbf{U}_\mathrm{t}\mathbf{H}_\mathrm{BR}\mathbf{w}_\mathrm{c}\right\|^2+\left\|\mathbf{U}_\mathrm{t}\mathbf{H}_\mathrm{BR}\mathbf{w}_\mathrm{b}\right\|^2)}{\sigma^2_\mathrm{e}}\Big)}_{f_5}\frac{P_0}{\ln2}.
\end{align}
Then, the MMSE method \cite{hua2020intelligent} is utilized to equivalently convert $f_1$, $f_2$, $f_3$, $f_4$ and $f_5$ as
\begin{align}
	&f_1=\max_{W_1>0, u_1}\ln(W_1)-W_1E_1(u_1, \mathbf{w}_\mathrm{b}, \mathbf{w}_\mathrm{c})+1,\label{eq_f1_trans}\\
	&f_2=\max_{W_2>0, \mathbf{u}_2}\ln(W_2)-W_2E_2(\mathbf{u}_2, \mathbf{w}_\mathrm{b}, \mathbf{w}_\mathrm{c})+1,\label{eq_f2_trans}\\
	&f_3=\max_{W_3>0, \mathbf{u}_3}\ln(W_3)-W_3E_3(\mathbf{u}_3, \mathbf{w}_\mathrm{b}, \mathbf{w}_\mathrm{c})+1,\label{eq_f3_trans}\\
	&f_4=\max_{W_4>0}\ln(W_4)-W_4E_4(\mathbf{w}_\mathrm{b}, \mathbf{w}_\mathrm{c})+1,\label{eq_f4_trans}\\
	&f_5=\max_{W_5>0}\ln(W_5)-W_5E_5(\mathbf{w}_\mathrm{b}, \mathbf{w}_\mathrm{c})+1,\label{eq_f5_trans}
\end{align}
where
\begin{itemize}
	\item $E_1(u_1, \mathbf{w}_\mathrm{b}, \mathbf{w}_\mathrm{c})=\left(u_1^H\mathbf{h}_\mathrm{rb}^H\mathbf{U}_\mathrm{r}\mathbf{H}_\mathrm{BR}\mathbf{w}_\mathrm{b}-1\right)\big(u_1^H\mathbf{h}_\mathrm{rb}^H\mathbf{U}_\mathrm{r}\\\mathbf{H}_\mathrm{BR}\mathbf{w}_\mathrm{b}-1\big)^H+u_1\left(\left|\mathbf{h}_\mathrm{rb}^H\mathbf{U}_\mathrm{r}\mathbf{H}_\mathrm{BR}\mathbf{w}_\mathrm{c}\right|^2+\sigma_\mathrm{b}^2\right)u_1^H\textcolor{blue}{,}$
	\vspace{2mm}
	\item $E_2(\mathbf{u}_2, \mathbf{w}_\mathrm{b}, \mathbf{w}_\mathrm{c})=\left(\sqrt{l_\mathrm{re}}\mathbf{u}_2^H\mathbf{U}_\mathrm{r}\mathbf{H}_\mathrm{BR}\mathbf{w}_\mathrm{c}-1\right)\big(\sqrt{l_\mathrm{re}}\mathbf{u}_2^H\\\mathbf{U}_\mathrm{r}\mathbf{H}_\mathrm{BR}\mathbf{w}_\mathrm{c}-1\big)^H+\sigma_\mathrm{e}^2\mathbf{u}_2^H\mathbf{u}_2\textcolor{blue}{,}$
		\vspace{2mm}
	\item $E_3(\mathbf{u}_3, \mathbf{w}_\mathrm{b}, \mathbf{w}_\mathrm{c})=\left(\sqrt{l_\mathrm{re}}\mathbf{u}_3^H\mathbf{U}_\mathrm{t}\mathbf{H}_\mathrm{BR}\mathbf{w}_\mathrm{c}-1\right)\big(\sqrt{l_\mathrm{re}}\mathbf{u}_3^H\\\mathbf{U}_\mathrm{t}\mathbf{H}_\mathrm{BR}\mathbf{w}_\mathrm{c}-1\big)^H+\sigma_\mathrm{e}^2\mathbf{u}_3^H\mathbf{u}_3\textcolor{blue}{,}$
		\vspace{2mm}
	\item $E_4(\mathbf{w}_\mathrm{b}, \mathbf{w}_\mathrm{c})=1+\frac{l_\mathrm{re}(\left\|\mathbf{U}_\mathrm{r}\mathbf{H}_\mathrm{BR}\mathbf{w}_\mathrm{c}\right\|^2+\left\|\mathbf{U}_\mathrm{r}\mathbf{H}_\mathrm{BR}\mathbf{w}_\mathrm{b}\right\|^2)}{\sigma^2_\mathrm{e}},$
	\vspace{2mm}
	\item$ E_5(\mathbf{w}_\mathrm{b},
	 \mathbf{w}_\mathrm{c})=1+\frac{l_\mathrm{re}(\left\|\mathbf{U}_\mathrm{t}\mathbf{H}_\mathrm{BR}\mathbf{w}_\mathrm{c}\right\|^2+\left\|\mathbf{U}_\mathrm{t}\mathbf{H}_\mathrm{BR}\mathbf{w}_\mathrm{b}\right\|^2)}{\sigma^2_\mathrm{e}}.$
	 	\vspace{2mm}
\end{itemize}
Here, $W_1$, $W_2$, $W_3$, $W_4$, $W_5$, $u_1$, $\mathbf{u}_2$ and $\mathbf{u}_3$ are the introduced auxiliary variables. Actually, the optimal auxiliary variables can be solved easily by setting the first-order partial derivative of the objective functions w.r.t. these variables to zero. \textcolor{blue}{This follows from the fact that the objective functions in equations \eqref{eq_f1_trans} - \eqref{eq_f5_trans} are convex w.r.t. these variables.} Thus, the optimal solutions can be given by
\begin{itemize}
\item  $u^\mathrm{opt}_1=\frac{\mathbf{h}_\mathrm{rb}^H\mathbf{U}_\mathrm{r}\mathbf{H}_\mathrm{BR}\mathbf{w}_\mathrm{b}}
    {\left|\mathbf{h}_\mathrm{rb}^H\mathbf{U}_\mathrm{r}\mathbf{H}_\mathrm{BR}\mathbf{w}_\mathrm{b}\right|^2
    +\left|\mathbf{h}_\mathrm{rb}^H\mathbf{U}_\mathrm{r}\mathbf{H}_\mathrm{BR}\mathbf{w}_\mathrm{c}\right|^2+\sigma_\mathrm{b}^2}$, \\ $W^\mathrm{opt}_1=E^{-1}_1,$
\vspace{2mm}
\item $\mathbf{u}^\mathrm{opt}_2=\left(l_\mathrm{re}\mathbf{U}_\mathrm{r}\mathbf{H}_\mathrm{BR}\mathbf{w}_\mathrm{c}\mathbf{w}_\mathrm{c}^H\mathbf{H}_\mathrm{BR}^H(\mathbf{U}_\mathrm{r})^H+\sigma_\mathrm{e}^2\mathbf{I}\right)^{-1}\sqrt{l_\mathrm{re}}$\\$\mathbf{U}_\mathrm{r}\mathbf{H}_\mathrm{BR}\mathbf{w}_\mathrm{c}, ~W^\mathrm{opt}_2=E_2^{-1},$
\vspace{2mm}
\item $\mathbf{u}^\mathrm{opt}_3=\left(l_\mathrm{re}\mathbf{U}_\mathrm{t}\mathbf{H}_\mathrm{BR}\mathbf{w}_\mathrm{c}\mathbf{w}_\mathrm{c}^H\mathbf{H}_\mathrm{BR}^H\mathbf{U}_\mathrm{t}^H+\sigma_\mathrm{e}^2\mathbf{I}\right)^{-1}\sqrt{l_\mathrm{re}}$\\$\mathbf{U}_\mathrm{t}\mathbf{H}_\mathrm{BR}\mathbf{w}_\mathrm{c},~ W^\mathrm{opt}_3=E_3^{-1},$
\vspace{2mm}
\item $W^\mathrm{opt}_4=E_4^{-1},~ W^\mathrm{opt}_5=E_5^{-1}.$
\vspace{2mm}
\end{itemize}
\textcolor{blue}{It is worth noting that we can easily demonstrate the equivalent relationship between the original logarithmic forms in \eqref{eq_Rb_trans} and the expressions in \eqref{eq_f1_trans} - \eqref{eq_f5_trans} by substituting the optimal auxiliary variables into \eqref{eq_f1_trans} - \eqref{eq_f5_trans}}.

According to the analysis above, we can obtain the concave lower bound of $f_1$, $f_2$, $f_3$, $f_4$ and $f_5$ in the $(t+1)$-th iteration with the given $W_1^{(t)}$, $W_2^{(t)}$, $W_3^{(t)}$, $W_4^{(t)}$, $W_5^{(t)}$, $u_1^{(t)}$, $\mathbf{u}_2^{(t)}$ and $\mathbf{u}_3^{(t)}$, which are expressed as
\begin{align}
&\widehat{f}_1(\mathbf{w}_\mathrm{b}, \mathbf{w}_\mathrm{c})=\ln(W_1^{(t)})-W_1^{(t)}E_1(u_1^{(t)}, \mathbf{w}_\mathrm{b}, \mathbf{w}_\mathrm{c})+1,\\
&\widehat{f}_2(\mathbf{w}_\mathrm{b}, \mathbf{w}_\mathrm{c})=\ln(W_2^{(t)})-W_2^{(t)}E_2(\mathbf{u}_2^{(t)}, \mathbf{w}_\mathrm{b}, \mathbf{w}_\mathrm{c})+1,\\
&\widehat{f}_3(\mathbf{w}_\mathrm{b}, \mathbf{w}_\mathrm{c})=\ln(W_3^{(t)})-W_3^{(t)}E_3(\mathbf{u}_3^{(t)}, \mathbf{w}_\mathrm{b}, \mathbf{w}_\mathrm{c})+1,\\
&\widehat{f}_4(\mathbf{w}_\mathrm{b}, \mathbf{w}_\mathrm{c})=\ln(W_4^{(t)})-W_4^{(t)}E_4(\mathbf{w}_\mathrm{b}, \mathbf{w}_\mathrm{c})+1,\\
&\widehat{f}_5(\mathbf{w}_\mathrm{b}, \mathbf{w}_\mathrm{c})=\ln(W_5^{(t)})-W_5^{(t)}E_5(\mathbf{w}_\mathrm{b}, \mathbf{w}_\mathrm{c})+1.
\end{align}
It is worth noting that $W_1^{(t)}$, $W_2^{(t)}$, $W_3^{(t)}$, $W_4^{(t)}$, $W_5^{(t)}$, $u_1^{(t)}$, $\mathbf{u}_2^{(t)}$ and $\mathbf{u}_3^{(t)}$ are calculated according to the obtained active beamforming solutions in the $t$-th iteration, i.e.,$\mathbf{w}_\mathrm{b}^{(t)}$ and $\mathbf{w}_\mathrm{c}^{(t)}$.
Thus, the concave lower bound of $\tilde{R}_\mathrm{b}^\mathrm{sa}$ in the $(t+1)$-th iteration can be expressed as
\begin{align}
	\widehat{R}_\mathrm{b}^\mathrm{sa}=\frac{1}{\ln2}\widehat{f}_1+\frac{P_1}{\ln2}\widehat{f}_2+\frac{P_0}{\ln2}\widehat{f}_3+\frac{P_1}{\ln2}\widehat{f}_4+\frac{P_0}{\ln2}\widehat{f}_5.
\end{align}

Similarly, in the $(t+1)$-th iteration, $R_\mathrm{c}$ can be equivalently transformed as
\begin{align}
	\tilde{R}_\mathrm{c}=\frac{1}{\ln2}\left(\ln (W_\mathrm{c}^{(t)})-W_\mathrm{c}^{(t)}E_\mathrm{c}(u_\mathrm{c}^{(t)},\mathbf{w}_\mathrm{b}, \mathbf{w}_\mathrm{c})+1\right),
\end{align}
where
\begin{itemize}
	\item $E_\mathrm{c}(u_\mathrm{c}^{(t)}, \mathbf{w}_\mathrm{b}, \mathbf{w}_\mathrm{c}) =\Big((u_\mathrm{c}^{(t)})^H\mathbf{h}_\mathrm{rc}^H\mathbf{U}_\mathrm{t}\mathbf{H}_\mathrm{BR}\mathbf{w}_\mathrm{c}-1\Big)\times$\\ $\Big((u_\mathrm{c}^{(t)})^H\mathbf{h}_\mathrm{rc}^H\mathbf{U}_\mathrm{t}\mathbf{H}_\mathrm{BR}\mathbf{w}_\mathrm{c}-1\Big)^H
+u_\mathrm{c}^{(t)}\Big(\left|\mathbf{h}_\mathrm{rc}^H\mathbf{U}_\mathrm{t}\mathbf{H}_\mathrm{BR}\mathbf{w}_\mathrm{b}\right|^2$\\
$+\sigma_\mathrm{c}^2\Big)\left(u_\mathrm{c}^{(t)}\right)^H$,
	\vspace{2mm}
\item $u_\mathrm{c}^{(t)}=\frac{\mathbf{h}_\mathrm{rc}^H\mathbf{U}_\mathrm{t}\mathbf{H}_\mathrm{BR}\mathbf{w}_\mathrm{c}^{(t)}}{\left|\mathbf{h}_\mathrm{rc}^H\mathbf{U}_\mathrm{t}\mathbf{H}_\mathrm{BR}\mathbf{w}_\mathrm{c}^{(t)}\right|^2+\left|\mathbf{h}_\mathrm{rc}^H\mathbf{U}_\mathrm{t}\mathbf{H}_\mathrm{BR}\mathbf{w}_\mathrm{b}^{(t)}\right|^2+\sigma_\mathrm{c}^2}$, \\ $W^{(t)}_\mathrm{c}=\left(E_\mathrm{c}^{(t)}\right)^{-1}.$
	\vspace{2mm}
\end{itemize}

Therefore, the subproblem \eqref{eq_active} can be reformulated as a convex optimization problem in the $(t+1)$-th iteration, which is given by
\begin{subequations}\label{eq_active_trans}
	\begin{align}
		&	\max_{\mathbf{w}_\mathrm{b}, \mathbf{w}_\mathrm{c}}~\omega_1\widehat{R}_\mathrm{b}^\mathrm{sa}+\omega_2\tilde{R}_\mathrm{c}\notag\\
		&\quad	\text{s.t.}~\mathbf{w}_\mathrm{b}^H\mathbf{w}_\mathrm{b}+\mathbf{w}_\mathrm{c}^H\mathbf{w}_\mathrm{c}\leq P_\mathrm{tmax}.\label{eq_active_trans_1}
	\end{align}
\end{subequations}
The Lagrange multiplier method \cite{boyd04} is further leveraged to solve the optimization problem \eqref{eq_active_trans} in this paper. Specifically, the Lagrange function of \eqref{eq_active_trans} can be expressed as
\begin{align}\label{eq_active_trans_La}
	\mathcal{L}(\mathbf{w}_\mathrm{b}, \mathbf{w}_\mathrm{c},\varrho)=&-\left(\omega_1\widehat{R}_\mathrm{b}^\mathrm{sa}+\omega_2\tilde{R}_\mathrm{c}\right)\notag\\
&+\varrho\Big(\mathbf{w}_\mathrm{b}^H\mathbf{w}_\mathrm{b}+\mathbf{w}_\mathrm{c}^H\mathbf{w}_\mathrm{c}
	-P_\mathrm{tmax}\Big),
\end{align}
where $\varrho$ denotes the Lagrange multiplier (dual variable) associated with the constraint \eqref{eq_active_trans_1}.
Note that the optimal solution of the Lagrange problem \eqref{eq_active_trans_La} for $\mathbf{w}_\mathrm{b}$ and $ \mathbf{w}_\mathrm{c}$ can be achieved by the following theorem.
\begin{theorem}
	The optimal $\mathbf{w}_\mathrm{b}$ and $ \mathbf{w}_\mathrm{c}$ of the problem \eqref{eq_active_trans_La} are given by
	\begin{align}\label{eq_active_opt_La}
			\mathbf{w}_\mathrm{b}^\mathrm{opt}(\varrho)=\mathbf{G}_1^{-1}\mathbf{G}_2, ~\mathbf{w}_\mathrm{c}^\mathrm{opt}(\varrho)=\widehat{\mathbf{G}}_1^{-1}\widehat{\mathbf{G}}_2,
	\end{align}
\begin{itemize}
	\item $\mathbf{G}_1=\varrho^\mathrm{opt}\mathbf{I}_{N_\mathrm{t}\times N_\mathrm{t}}+\frac{\omega_2W_\mathrm{c}^{(t)}\left|u_\mathrm{c}^{(t)}\right|^2}{\ln2}\mathbf{A}+\frac{\omega_1W_1^{(t)}\left|u_1^{(t)}\right|^2}{\ln2}\mathbf{B}\\+\frac{\omega_1l_\mathrm{re}P_1W_4^{(t)}}{\sigma_\mathrm{e}^2\ln2}\mathbf{C}_\mathrm{r}+\frac{\omega_1l_\mathrm{re}P_0W_5^{(t)}}{\sigma_\mathrm{e}^2\ln2}\mathbf{C}_\mathrm{t}$,
	\vspace{3mm}
	\item $\mathbf{G}_2=\frac{\omega_1W_1^{(t)}}{\ln2}\mathbf{H}_\mathrm{BR}^H(\mathbf{U}^\mathrm
	{ES}_\mathrm{r})^H\mathbf{h}_\mathrm{rb}u_1^{(t)}$,
	\vspace{3mm}
	\item $\widehat{\mathbf{G}}_1=\varrho^\mathrm{opt}\mathbf{I}_{N_\mathrm{t}\times N_\mathrm{t}}+\frac{\omega_2W_\mathrm{c}^{(t)}}{\ln2}\left|u_\mathrm{c}^{(t)}\right|^2\mathbf{A}+\frac{\omega_1W_1^{(t)}}{\ln2}\left|u_1^{(t)}\right|^2\times\\\mathbf{B}+\frac{\omega_1P_1W_2^{(t)}l_\mathrm{re}}{\ln2}\mathbf{D}+\frac{\omega_1P_0W_3^{(t)}l_\mathrm{re}}{\ln2}\tilde{\mathbf{D}}+\frac{\omega_1l_\mathrm{re}P_1W_4^{(t)}}{\sigma_\mathrm{e}^2\ln2}\mathbf{C}_\mathrm{r}\\+\frac{\omega_1l_\mathrm{re}P_0W_5^{(t)}}{\sigma_\mathrm{e}^2\ln2}\mathbf{C}_\mathrm{t}$,
	\vspace{3mm}
	\item $\widehat{\mathbf{G}}_2=\frac{\omega_2W_\mathrm{c}^{(t)}}{\ln2}\mathbf{H}_\mathrm{BR}^H\mathbf{U}_\mathrm{t}^H\mathbf{h}_\mathrm{rc}u_\mathrm{c}^{(t)}+\frac{\omega_1P_1W_2^{(t)}\sqrt{l_\mathrm{re}}}{\ln2}\times\\
\mathbf{H}_\mathrm{BR}^H\mathbf{U}_\mathrm{r}^H\mathbf{u}_2^{(t)}+\frac{\omega_1P_0W_3^{(t)}\sqrt{l_\mathrm{re}}}{\ln2}\mathbf{H}_\mathrm{BR}^H\mathbf{U}_\mathrm{t}^H\mathbf{u}_3^{(t)}$,
	\vspace{3mm}
	\item $\mathbf{A}=\mathbf{H}_\mathrm{BR}^H\mathbf{U}_\mathrm{t}^H\mathbf{h}_\mathrm{rc}\mathbf{h}_\mathrm{rc}^H\mathbf{U}_\mathrm{t}\mathbf{H}_\mathrm{BR},~ \mathbf{C}_\mathrm{r}=\mathbf{H}_\mathrm{BR}^H\mathbf{U}_\mathrm{r}^H\mathbf{U}_\mathrm{r}\mathbf{H}_\mathrm{BR},$
		\vspace{3mm}
	 \item $\mathbf{B}=\mathbf{H}_\mathrm{BR}^H\mathbf{U}_\mathrm{r}^H\mathbf{h}_\mathrm{rb}\mathbf{h}_\mathrm{rb}^H\mathbf{U}_\mathrm{r}\mathbf{H}_\mathrm{BR}, ~ \mathbf{C}_\mathrm{t}=\mathbf{H}_\mathrm{BR}^H\mathbf{U}_\mathrm{r}^H\mathbf{U}_\mathrm{r}$,
	\vspace{3mm}
	\item $\mathbf{D}=\mathbf{H}_\mathrm{BR}^H\mathbf{U}_\mathrm{r}^H\mathbf{u}_2^{(t)}\left(\mathbf{u}_2^{(t)}\right)^H\mathbf{U}_\mathrm{r}\mathbf{H}_\mathrm{BR}, ~\tilde{\mathbf{D}}=\mathbf{H}_\mathrm{BR}^H\mathbf{U}_\mathrm{t}^H\mathbf{u}_3^{(t)}$\\$\left(\mathbf{u}_3^{(t)}\right)^H\mathbf{U}_\mathrm{t}\mathbf{H}_\mathrm{BR}$.
		\vspace{3mm}
\end{itemize}
\begin{proof}
	The proof is provided in Appendix \ref{app_1}.
\end{proof}
\end{theorem}
\begin{remark}
The optimal results of active beamforming, as given by \eqref{eq_active_opt_La}, reveal a close connection between the optimal active beamforming and the dual variable $\varrho$. For any given $\varrho$, we can directly calculate the optimal active beamforming in closed-form. Therefore, to achieve the optimal active beamforming of the original problem \eqref{eq_active_trans}, it is crucial to solve its dual problem and subsequently determine the optimal dual variable, since this is a key step in achieving the optimal active beamforming for the original problem \cite{boyd04}.
\end{remark}

The dual problem of the problem \eqref{eq_active_trans} is given by
\begin{align}\label{eq_dual_pro}
	\max_{\varrho\geq0}g(\varrho),
\end{align}
where $g(\varrho)$ represents the dual function expressed as
\begin{align}
		g(\varrho)=\min_{\mathbf{w}_\mathrm{b}, \mathbf{w}_\mathrm{c}}\mathcal{L}(\mathbf{w}_\mathrm{b}, \mathbf{w}_\mathrm{c},\varrho).
\end{align}
The subgradient method \cite{hu2019uav} is utilized to address the dual problem and the updating rule of the dual variable $\varrho$ in the $(l+1)$-th inner loop iteration can be described as
\begin{align}\label{eq_dual_var}
	\varrho^{(l+1)}=&\varrho^{(l)}+\varpi\Big(\left\|\mathbf{w}^\mathrm{opt}_\mathrm{b}(\varrho^{(l)})\right\|^2+\left\|\mathbf{w}^\mathrm{opt}_\mathrm{c}(\varrho^{(l)})\right\|^2\notag\\
	&~~~~~~~~~~~~-P_\mathrm{tmax}\Big),
\end{align}
where  $\varpi$ represents the positive increment used to update $\varrho^{(l)}$ in the algorithm. The detailed processes of solving the active beamforming variables are summarized in Algorithm 1.
 \begin{center}
	\begin{tabular}{p{8.5cm}}
		\toprule[2pt]
		\textbf{Algorithm 1:}  Subgradient Method for Solving Dual Problem \eqref{eq_dual_pro}   \\
		\midrule[1pt]
		1: Initialize $\varrho^{(0)}\geq 0$ and $\varpi>0$; Set the iteration index\\
		\quad~ $l$ = 0.\\
		2: \textbf{Repeat} \\
		3: \quad Calculate $\mathbf{w}^\mathrm{opt}_\mathrm{b}(\varrho^{(l)})$ and $\mathbf{w}^\mathrm{opt}_\mathrm{c}(\varrho^{(l)})$ through \eqref{eq_active_opt_La}\\\qquad~based on $\varrho^{(l)}$.\\
		4: \quad Update $\varrho^{(l+1)}$ utilizing \eqref{eq_dual_var}. \\
		5: \quad Let $l=l+1$.\\
		6: \textbf{Until} the fractional increase of \eqref{eq_dual_pro} is smaller than
		a\\\quad~predefined threshold \\
		7: \textbf{Output:} $\mathbf{w}^\mathrm{opt}_\mathrm{b}$ and $\mathbf{w}^\mathrm{opt}_\mathrm{c}$\\
		\bottomrule[2pt]
	\end{tabular}
\end{center}

\vspace{2mm}
\subsubsection{Passive beamforming design} This section will design the passive beamforming variables, i.e., $\mathbf{U}_\mathrm{r}$ and $\mathbf{U}_\mathrm{t}$, according to the obtained $\mathbf{w}_\mathrm{b}$ and $\mathbf{w}_\mathrm{c}$. The initial optimization issue can be streamlined as
\begin{subequations}\label{eq_passive}
	\begin{align}
		&	\max_{\mathbf{U}_\mathrm{r},\mathbf{U}_\mathrm{t}}~\omega_1\tilde{R}_\mathrm{b}^\mathrm{sa}+\omega_2R_\mathrm{c}\notag\\
		&\quad	\text{s.t.}~~~\beta_{m}^{\mathrm{r}}, \beta_{m}^{\mathrm{t}}\in \mathcal{B}, ~\forall m\in\mathcal{M},\label{eq_passive_1} \\
		&\qquad\quad\beta_{m}^{\mathrm{r}}+\beta_{m}^{\mathrm{t}}=1,~ \forall m\in\mathcal{M},\label{eq_passive_2} \\
		&\qquad\quad \phi_m^{\mathrm{r}}, \phi_m^{\mathrm{r}}\in\{\frac{2\pi}{2^{\lambda}},2\frac{2\pi}{2^{\lambda} }, \cdots, 2^{\lambda}\frac{2\pi}{2^{\lambda}}\}.\label{eq_passive_3}
	\end{align}
\end{subequations}
Actually, addressing this subproblem presents a challenge due to the non-convexity of the objective function, the equal constraint on the amplitude of the STAR-RIS, and the discrete phase-shifts involved. To tackle this challenge, the CEO method \cite{rubinstein2004cross} is utilized to design the passive beamforming variables. Different form the conventional convex optimization algorithms, the CEO approach leverages the probability-based learning technique to solve complex problems. Its fundamental idea behind handling the complex optimization problem can be summarized as: (\romannumeral 1) Designing the appropriate sampling distribution characterized by the tilting parameters for different types of variables; (\romannumeral 2) Generating prospective solutions from the sample distribution and assessing these solutions through the computation of the relevant objective function value; (\romannumeral 3)  Adjusting the tilting parameters of the sampling distribution by optimizing the cross-entropy between the existing distribution and the benchmark distribution established from these notably efficacious solutions; (\romannumeral 4)  Iterating through new viable samples using the revised tilting parameters until the discrepancy of the objective function values between consecutive iterations falls a predefined threshold. The key steps are presented below.
\paragraph{Sampling distribution design for the discrete phase-shifts of the STAR-RIS}
In this section, the sampling distribution of the phase-shifts will be carefully structured to produce $2M$ phase-shifts (comprising reflection and transmission phases) within the set $\mathcal{T}\triangleq\{\varphi_{q}=q\frac{2\pi}{2^{\lambda}}|q\in\mathcal{Q}\triangleq\{1,\cdots, Q=2^{\lambda}\}$. Specifically, let $\boldsymbol{\phi}=\{\phi_p\}_{p=1}^{2M}\triangleq\{\{\phi_m^{\mathrm{r}}\}_{m=1}^M, \{\phi_m^{\mathrm{t}}\}_{m=1}^M\}$. It is worth noting that any $\phi_p$ can be mapped into the specific reflection or transmission phase-shifts easily. To generate the feasible phase-shifts, we can sample each $\phi_p,~ p\in\mathcal{P}\triangleq\{1, \cdots, p, \cdots, 2M\}$, in parallel and independently utilizing the acquired or predetermined discrete probability distribution $\{P_{p1}, \cdots, P_{pq}, \cdots, P_{pQ}\}$ , where $P_{pq}$ denotes the probability that $\phi_p$ opts the $q$-th entry in the set $\mathcal{T}$, and satisfies $\sum_{q=1}^{Q}P_{pq}=1$. Thus, the joint sampling distribution of generating $\boldsymbol{\phi}$ can be given by
\begin{align}
    \mathscr{G}(\boldsymbol{\phi}; \mathbf{P})=\prod_{p=1}^{2M}\sum_{q=1}^{Q}P_{pq}\mathbbm{1}_{\{\mathcal{I}_p(\boldsymbol{\phi})=\varphi_q\}},
\end{align}
where matrix $\mathbf{P}=[P_{pq}]\in\mathbb{R}^{P\times Q}$. $\mathcal{I}_p(\boldsymbol{\phi})$ denotes the $p$-th element in vector $\boldsymbol{\phi}$. $\mathbbm{1}_{\{\cdot\}}$ is the indicator function to a event.

\textcolor{blue}{Next, we will elaborate on the process of generating the discrete phase shift for the $ p $-th element based on its sampling distribution. Specifically, we begin by sampling a random number $a$ from a uniform distribution over the interval $[0,1]$. Then, leveraging this sampled value, we determine the discrete phase shift of the $p$-th element according to the following rule:
\begin{align}\label{eq_sampling_rule}
	\phi_p =\begin{cases}
		\frac{2\pi}{2^{\lambda}},& a\leq P_{p1},\\
		2\frac{2\pi}{2^{\lambda}},& P_{p1}\leq a\leq P_{p1}+P_{p2},\\
		~~\vdots, & ~~~\vdots\\	
		Q\frac{2\pi}{2^{\lambda}},& \sum_{q=1}^{Q-1}P_{pq}\leq a\leq \sum_{q=1}^{Q}P_{pq}=1.\\
	\end{cases}
\end{align}
Based on this generating rule, we observe that the probability for the random number
$a$ to fall within the interval $[\sum_{q=1}^{n-1}P_{pq}, ~\sum_{q=1}^{n}P_{pq}]$ is equal to the probability that $\phi_p =n\frac{2\pi}{2^{\lambda}}$. In other words, we have
\begin{align}
	\operatorname{Pr}\left\{\phi_p =n\frac{2\pi}{2^{\lambda}}\right\} &=\operatorname{Pr}\left\{\sum_{q=1}^{n-1}P_{pq}\leq a\leq \sum_{q=1}^{n}P_{pq}\right\}\notag\\
	&\stackrel{(b)}{=}\sum_{q=1}^{n}P_{pq}-\sum_{q=1}^{n-1}P_{pq}= P_{pn},
\end{align}
where $(b)$ is due to the equal probability distribution of the uniform distribution. This result confirms that, by using the generating rule specified in \eqref{eq_sampling_rule}, the generated $\phi_p$ is guaranteed to follow its intended sampling distribution. By applying this procedure independently to each element, the overall system achieves the desired phase-shift configuration as dictated by the joint sampling distribution}

\paragraph{Sampling distribution design for the continuous amplitudes of the STAR-RIS}
Selecting a suitable sampling distribution is crucial for generating continuous variables, which differs from discrete variables. A typical selection for the sampling distribution of generating the continuous variables is the Gaussian distribution \cite{xiao2024frequency,chen2024designing}. \textcolor{blue}{The reasons are as follows: (\romannumeral 1) The specific distribution of the amplitude coefficients for each element may be uncertain; however, in practical engineering scenarios, many random variables can be approximated by Gaussian distributions, making the use of a Gaussian distribution more suitable for real-world applications. (\romannumeral 2) Gaussian distributions possess the maximum entropy property, meaning that given a fixed mean and variance, the Gaussian distribution has the highest entropy. This indicates that in the absence of complete information, a Gaussian distribution provides the most unbiased and reasonable choice for characterizing the behavior of a random variable. (\romannumeral 3) By appropriately designing the mean and variance of the Gaussian distributions, it is possible to effectively control the range of sampled values and concentrate the probability mass within a specific interval. This characteristic is beneficial for iteratively optimizing the amplitude coefficients of the elements.
}
 Thus, the following joint sampling distribution is considered to sample the amplitude coefficients of the STAR-RIS, which is given by
\begin{align}\label{eq_sampling_ampli}
	\mathscr{N}(\boldsymbol{\beta}^{\mathrm{r}}; \boldsymbol{\mu}; \boldsymbol{\sigma})=\prod_{m=1}^{M}\frac{1}{\sqrt{2\pi}\sigma_{m}}e^{-\frac{(\mathcal{I}_m(\boldsymbol{\beta}^{\mathrm{r}})-\mu_{m})^2}{2\sigma_{m}^2}},
\end{align}
where $\boldsymbol{\sigma}=\{\sigma_{m}\}_{m=1}^M$ and $\boldsymbol{\mu}=\{\mu_{m}\}_{m=1}^M$ respectively represent the standard deviation and mean vector, where $\sigma_{m}$ and $\mu_{m}$ are key parameters to generate the amplitude of the $m$-th element. Also,  $\boldsymbol{\beta}^{\mathrm{r}}\triangleq\{\beta_{1}^{\mathrm{r}}, \cdots, \beta_{m}^{\mathrm{r}}, \cdots, \beta_{M}^{\mathrm{r}}\}$. Note that the sampling distribution \eqref{eq_sampling_ampli} is  solely capable of producing the amplitudes of the reflection coefficients. In fact, the amplitudes of the transmission coefficients can be directly calculated using the constraints \eqref{eq_passive_2}, i.e., $\beta_{m}^{\mathrm{t}}=1-\beta_{m}^{\mathrm{r}}$. It is possible for the sampling distribution to yield the amplitudes falling out the range $(0,1]$. To address this issue, we can employ the subsequent techniques to ensure the generated amplitudes are viable: (\romannumeral 1) Removing the infeasible amplitudes and randomly selecting the amplitudes in the required range to replace them; (\romannumeral 2) Iteratively generating alternative value to each infeasible amplitude until a feasible amplitude is achieved.

On the basis of the analysis above, the joint sampling distribution for tackling the optimization problem \eqref{eq_passive} can be derived as
\begin{align}\label{eq_joint_distri}
	\mathscr{F}(\boldsymbol{\Xi};\boldsymbol{\Upsilon})=\mathscr{G}\times	\mathscr{N},
\end{align}
where $\boldsymbol{\Xi}=\{\boldsymbol{\phi}, \boldsymbol{\beta}^{\mathrm{r}}\}$,  $\boldsymbol{\Upsilon}=\{\mathbf{P}, \boldsymbol{\mu}, \boldsymbol{\sigma}\}$.  It is worth noting that the sampling distribution presented in \eqref{eq_joint_distri} will be leveraged to produce the feasible solutions of the problem \eqref{eq_passive}.
\paragraph{Updating formulas for tilting parameters}
In this section, the tilting parameters, i.e., $\boldsymbol{\Upsilon}$, will be updated based on the distribution derived from the top-tier feasible solutions. Specifically, we first utilize the sampling distribution in \eqref{eq_joint_distri} to yield $K$ feasible solutions $\{\boldsymbol{\Xi}_k\}_{k=1}^K$ and compute the objective function value of each $\boldsymbol{\Xi}_k$, denoted as $R(\boldsymbol{\Xi}_k)$. Then, the viable solutions will be prioritized in a systematic manner based on the objective function values to acquire high-performing samples. Specifically, we re-denote the solution with the $k$-th largest objective value as $\boldsymbol{\Xi}_{[k]}$, which means $R(\boldsymbol{\Xi}_{[1]})\geq R(\boldsymbol{\Xi}_{[2]})\geq\cdots\geq R(\boldsymbol{\Xi}_{[k]})\geq\cdots\geq R(\boldsymbol{\Xi}_{[K]})$. In addition,the first $K^\mathrm{elite}=\eta K$ samples are selected to establish the elite set $\mathcal{E}$ based on the order above, where $\eta\in(0, 1)$ is the proportion of solutions within the elite set relative to the total number of samples.

Note that, the tilting parameters are adjusted by minimizing the cross-entropy between the current distribution and the new distribution derived from the elite samples in the CEO framework. As shown in \cite{rubinstein2004cross}, the cross-entropy minimization optimization problem can be equivalently converted as
\begin{subequations}\label{eq_entropy}
	\begin{align}
		&	\max_{\boldsymbol{\Upsilon}}~\frac{1}{K}\sum_{k=1}^{K^\mathrm{elite}}\ln	\mathscr{F}(\boldsymbol{\Xi}_{[k]};\boldsymbol{\Upsilon})\notag\\
		&\quad	\text{s.t.}~\sum_{q=1}^{Q}P_{pq}=1. \label{eq_entropy_1}
	\end{align}
\end{subequations}
Here, we provide the following Theorem \ref{th2} which gives the close-form solution of the problem \eqref{eq_entropy}.
\begin{theorem}\label{th2}
	The updating formulas of the tilting parameters can be given by
	\begin{align}
		&P_{pq}=-\frac{\sum_{k=1}^{K^\mathrm{elite}}\mathbbm{1}_{\{\mathcal{I}_p(\boldsymbol{\phi}_{[k]})=\varphi_{q}\}}}{K^\mathrm{elite}}, ~\forall p\in\mathcal{P}, ~q\in\mathcal{Q}\label{eq_P_update}\\
		&\sigma_{m}=\sqrt{\frac{\sum_{k=1}^{K^\mathrm{elite}}\Big(\mathcal{I}_m(\boldsymbol{\beta}_{[k]}^\mathrm{r})-\mu_m\Big)^2}{K^\mathrm{elite}}}, ~\forall m\in\mathcal{M},\label{eq_sigma_update}\\
		&\mu_{m}=\frac{\sum_{k=1}^{K^\mathrm{elite}}\mathcal{I}_m(\boldsymbol{\beta}_{[k]}^\mathrm{r})}{K^\mathrm{elite}},  ~\forall m\in\mathcal{M}.\label{eq_mu_update}
	\end{align}
\begin{proof}
	The proof is provided in Appendix \ref{app_2}.
\end{proof}
\end{theorem}

A smoothing technique is integrated into the fine-tuning of tilting parameters, commonly employed in reinforcement learning, in order to mitigate the risk of converging towards local optimal solutions. Therefore, the tilting parameters in the $(i+1)$-th inner loop iteration are adjusted based on the subsequent approach:
\begin{align}
	&P_{pq}^{(i+1)}\leftarrow\chi P_{pq}^{(i+1)}+(1-\chi)P_{pq}^{(i)},\label{eq_P_smmoth}\\
	&\sigma_{m}^{(i+1)}\leftarrow\chi \sigma_{m}^{(i+1)}+(1-\chi)\sigma_{m}^{(i)},\label{eq_sigma_smmoth}\\
	&\mu_{m}^{(i+1)}\leftarrow\chi \mu_{m}^{(i+1)}+(1-\chi)\mu_{m}^{(i)},\label{eq_mu_smmoth}
\end{align}
where $\chi\in(0, 1)$ is the smoothing factor.

Algorithm 2 summarizes the comprehensive procedures of CEO method to obtain the optimal passive beamforming variables, which is presented as following:
\begin{center}
	\begin{tabular}{p{8.5cm}}
		\toprule[2pt]
		\textbf{Algorithm 2:}  CEO Method for Solving Optimization Problem \eqref{eq_passive}   \\
		\midrule[1pt]
		1: Initialize $P^{(0)}_{pq}=1/Q, ~\forall p\in\mathcal{P}, ~q\in\mathcal{Q}$, $\mu^{(0)}_{m}=1/2,~ \forall m$\\\quad $\in\mathcal{M}$, $\sigma^{(0)}_{m}=1,~ \forall m\in\mathcal{M}$, $R^\mathrm{best}$ and $\boldsymbol{\Xi}^\mathrm{best}$; Set the\\\quad iteration index $i$ = 0.\\
		2: \textbf{Repeat} \\
		3:  \quad Generate $K$ feasible solutions $\{\boldsymbol{\Xi}_k^{(i)}\}_{k=1}^K$ based on the \\\qquad distribution sampling $\mathscr{F}(\cdot;\boldsymbol{\Upsilon}^{(i)})$ and calculate their\\\qquad objective function values $\{R(\boldsymbol{\Xi}_k^{(i)})\}_{k=1}^K$. \\
		4: \quad Sort $\{\boldsymbol{\Xi}_k^{(i)}\}_{k=1}^K$ based on $\{R(\boldsymbol{\Xi}_k^{(i)})\}_{k=1}^K$  in descend\\\qquad order, i.e., $R(\boldsymbol{\Xi}_{[1]}^{(i)})\geq\cdots\geq R(\boldsymbol{\Xi}_{[k]}^{(i)})\geq\cdots\geq R(\boldsymbol{\Xi}_{[K]}^{(i)})$\\\qquad and choose the first $K^\mathrm{elite}=\eta K$ solutions to establish\\\qquad the elite set $\mathcal{E}$. \\
		5: \quad Update the tilting parameters $P_{pq}^{(i+1)}$, $\sigma_{m}^{(i+1)}$ and $\mu_{m}^{(i+1)}$\\\qquad with the elite solutions according to \eqref{eq_P_update}-\eqref{eq_mu_update}. \\
		6: \quad Implement the smoothing process for the tilting pa-\\\qquad rameters according to \eqref{eq_P_smmoth}-\eqref{eq_mu_smmoth} and achieve $\boldsymbol{\Upsilon}^{(i+1)}$.\\
		7: \quad if $R(\boldsymbol{\Xi}_{[1]}^{(i)})\geq R^\mathrm{best}$\\
		8: \qquad $R^\mathrm{best}\leftarrow R(\boldsymbol{\Xi}_{[1]}^{(i)})$;~ $\boldsymbol{\Xi}^\mathrm{best}\leftarrow \boldsymbol{\Xi}_{[1]}^{(i)}$.\\
		9: \quad end\\
		10:\quad Let $i=i+1$. \\
		11: \textbf{Until} the difference in the optimal objective function\\\quad~ values between two successive iterations is smaller than\\\quad~ a predefined threshold. \\
		12: \textbf{Output:} $\boldsymbol{\Xi}^\mathrm{best}$ and then determine the optimal $\mathbf{U}_\mathrm{r}$\\\quad~ and $\mathbf{U}_\mathrm{t}$.\\
		\bottomrule[2pt]
	\end{tabular}
\end{center}
\subsection{Proposed Algorithm \& Analysis on Complexity and Convergence}
Algorithm 3 outlines the comprehensive process of the proposed method for tackling the original optimization problem \eqref{eq_oringi_opti}. It iteratively addresses the active beamforming and passive beamforming problems until the discrepancy in the objective function values from one iteration to the next falls below a predetermined threshold.

The computational complexity of the algorithm under consideration is primarily determined by addressing two key subproblems, i.e, the active beamforming and passive beamforming problems. Algorithm 1 is employed to effectively address the first subproblem, where the primary computational challenge arises from computing the subgradient while solving the dual problem. The computational complexity is expressed as $\mathcal{O}(L_1N_\mathrm{t}^3)$, where $L_1$ represents the quantity of iterations required to determine the optimal dual variable $\varrho$. Algorithm 2 addresses the passive beamforming problem with discrete phase-shift variables. The main computational challenge comes from the necessity to assess the objective function values of $K$ viable solutions, resulting in a computational complexity of $\mathcal{O}(L_2K)$, where $L_2$ denotes the number of iterations executed by Algorithm 2. Thus, the total computing complexity of the proposed algorithm can be calculated as $\mathcal{O}(L_\mathrm{tol}(L_1N_\mathrm{t}^3+L_2K))$, where $L_\mathrm{tol}$ is the total number of iterations for Algorithm 3.
\begin{center}
	\begin{tabular}{p{8.5cm}}
		\toprule[2pt]
		\textbf{Algorithm 3:}  The Proposed Algorithm for Solving the Original Optimization Problem \eqref{eq_oringi_opti}   \\
		\midrule[1pt]
		1: Initialize feasible $(\mathbf{w}_\mathrm{b}^{(0)}, \mathbf{w}_\mathrm{c}^{(0)},\mathbf{U}_\mathrm{r}^{(0)},\mathbf{U}_\mathrm{t}^{(0)})$; Set the itera-\\\quad tion index $t$ = 0.\\
		2: \textbf{Repeat} \\
		3: \quad Solve the subproblem \eqref{eq_active} utilizing Algorithm 1 with\\\qquad given $(\mathbf{U}_\mathrm{r}^{(t)},\mathbf{U}_\mathrm{t}^{(t)})$ and update $(\mathbf{w}_\mathrm{b}^{(t+1)}, \mathbf{w}_\mathrm{c}^{(t+1)})$ with\\\qquad the acquired solutions.\\
		4: \quad Solve the subproblem \eqref{eq_passive} utilizing Algorithm 2 with\\\qquad given $(\mathbf{w}_\mathrm{b}^{(t+1)}, \mathbf{w}_\mathrm{c}^{(t+1)})$ and update $(\mathbf{U}_\mathrm{r}^{(t+1)},\mathbf{U}_\mathrm{t}^{(t+1)})$\\\qquad with the acquired solutions.\\
		5: \textbf{Until} the difference of the objective function values \\\quad between two successive iterations is smaller than a pre-\\\quad defined threshold; Let $t=t+1$.\\
		6: \textbf{Output:} the optimal $\mathbf{w}_\mathrm{b}$,  $\mathbf{w}_\mathrm{c}$, $\mathbf{U}_\mathrm{r}$ and $\mathbf{U}_\mathrm{t}$.\\
		\bottomrule[2pt]
	\end{tabular}
\end{center}
\begin{remark}
	To achieve the desired performance, \cite{rubinstein2004cross} recommends that $K$ should satisfy $K=\omega N_\mathrm{var}$, where $\omega\in[4, 10]$, $N_\mathrm{var}$ represents the total number of variables. We can find that Algorithm 2 aims to optimize a total of $3M$ variables. Hence, the total computing complexity of the proposed algorithm can be further expressed as $\mathcal{O}(L_\mathrm{tol}(L_1N_\mathrm{t}^3+3L_2\omega M))$. It is worth highlighting that the computational complexity of Algorithm 2 founded on the CEO framework is significantly lower than that of numerous existing optimization algorithms. Therefore, the suggested algorithm demonstrates a notable benefit and potential in addressing optimization challenges within practical communication systems assisted by RIS.
\end{remark}

The proposed algorithm utilizes the alternating strategy to solve the non-convex optimization problem, thus, it can ensure that the current objective function value is not lower than that of the previous iteration during the iterative process. The validation of the proposed algorithm's convergence and effectiveness will be extensively examined through the simulation findings in Section \ref{sec:S5}. \textcolor{blue}{Although the proposed algorithm is initially designed based on the ES protocol, it can be readily extended to accommodate the mode switching (MS) and time switching (TS) protocols of STAR-RIS. This adaptability stems from the algorithm's capability to effectively address the key challenges associated with these protocols. Specifically, in the MS protocol, each STAR-RIS element operates in either reflection or transmission mode with binary amplitude control, which will introduce the discrete variable design challenge. Similarly, the TS protocol introduces the challenge of time allocation optimization, requiring a precise balance between reflected and transmitted duration. Note that, the proposed algorithm can jointly optimize both discrete and continuous variables and show a superior performance, ensuring its effectiveness for the other two protocols.}

\section{Numerical Simulation}\label{sec:S5}
In this section, the effectiveness of the proposed STAR-RIS-assisted PLS scheme is verified by comparing the numerical simulation results with the following benchmark schemes:

\textbf{1) SDR scheme:} The semidefinite relaxation (SDR) method is utilized to design the passive beamforming variables of the STAR-RIS. Note that the practical implement for the phase control of the STAR-RIS is considered in this paper, which means the STAR-RIS can only provide finite discrete phase-shifts for the incident signals. In this scheme, we first optimize the reflection and transmission coefficients taking into account continuous phase-shifts. Then, we can obtain the discrete phase-shifts by the following operation:
	\begin{align}
		\phi^{\mathrm{opt},{Y}}_m=&\arg\min_{\phi^{Y}_m\in\mathcal{T}}\left|\phi^{Y}_m-\phi_m^{\mathrm{c}, Y}\right|, \forall m\in\mathcal{M},\notag\\& Y\in\{\mathrm{r, t}\},
	\end{align}
	where $\phi_m^{\mathrm{c}, Y}$ denotes the continuous reflection or transmission phase-shift of the $m$-th element solved by the SDR method.

 \textbf{2) ZF scheme:} The active beamforming variables, i.e., $\mathbf{w}_\mathrm{b}$ and $\mathbf{w}_\mathrm{c}$ are designed by the zero-forcing (ZF) approach in this scheme. Specifically, the corresponding  $\mathbf{w}_\mathrm{b}$ and $\mathbf{w}_\mathrm{c}$ can be derived as
	\begin{align}
\mathbf{w}_\mathrm{b}^\mathrm{opt}=\sqrt{P_\mathrm{b}}\frac{\mathbf{W}(:,1)}{\left|\mathbf{W}(:,1)\right|}, ~ \mathbf{w}_\mathrm{c}^\mathrm{opt}=\sqrt{P_\mathrm{c}}\frac{\mathbf{W}(:,2)}{\left|\mathbf{W}(:,2)\right|},
	\end{align}
	where $\mathbf{W}=(\mathbf{H}\mathbf{H}^H)^{-1}\mathbf{H}^H$, and $\mathbf{H}=[(\mathbf{h}_\mathrm{rb}^H\boldsymbol{\Theta}_\mathrm{r}\mathbf{H}_\mathrm{BR})^T,$  $(\mathbf{h}_\mathrm{rc}^H\boldsymbol{\Theta}_\mathrm{t}\mathbf{H}_\mathrm{BR})^T]^T$. In addition,  $P_\mathrm{b}$ and $P_\mathrm{c}$ respectively are the allocated transmitting power for Bob and Carol, which require carefully design.

  \textbf{3) RIS-aided scheme:} To achieve the full-space coverage, two adjacent traditional RISs are leveraged to replace the STAR-RIS in this scheme, where one is the reflection-only RIS while the other one is the transmission-only RIS. In order to guarantee the fairness, each RIS equips with $\frac{M}{2}$ elements.

Additionally, the parameters implemented the simulation are listed in Table \ref{tab:table1}.
\begin{table}[h!]
	\renewcommand\arraystretch{1.5}
	\centering
	\caption{Parameters Setting}
	\label{tab:table1}
	\begin{tabular}{|M{3.8cm}|M{4cm}|}
		\hline		
		\textbf{\normalsize{Parameters}}& \textbf{\normalsize{Symbol and Value}}\\
		\hline
		\normalsize{Reference power gain}&\normalsize{$\rho=-20$ dB}   \\
		\hline
		\normalsize{Path-loss exponent }& \normalsize{$\alpha=2.6$}\\
		\hline
		\normalsize{Distance between the STAR-RIS and BS/users}& \normalsize{$d_\mathrm{BR}=400$ m, $d_\mathrm{rb}=75$ m, $d_\mathrm{rc}=100$ m}\\
		\hline
		\normalsize{Noise power}& \normalsize{$\sigma_\mathrm{b}^2=\sigma_\mathrm{c}^2=\sigma_\mathrm{e}^2=-110$ dBm} \\
		\hline
		\normalsize{Parameters associated with Algorithm 2}& \normalsize{$\omega=4$, $\eta=0.1$, $\chi=0.55$}\\
		\hline
	\end{tabular}
\end{table}

\begin{figure}[ht]
	\centering
	\includegraphics[scale=0.45]{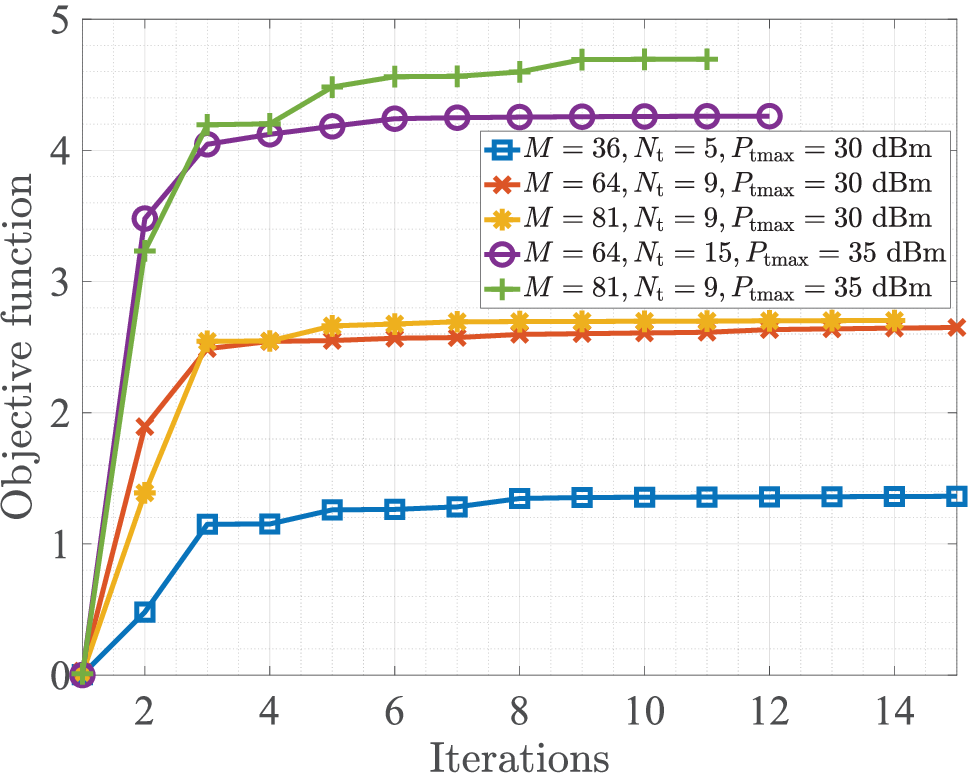}\\
	\caption{Iterative curves of the proposed algorithm considering $P_1=P_0=0.5$, $\omega_1=\omega_2=0.5$, $\lambda=2$ and different $M$, $N_\mathrm{t}$ and $P_\mathrm{tmax}$.}\label{fig:obj_vs_iteration}
\end{figure}
Fig. \ref{fig:obj_vs_iteration} shows the iterative curves of the proposed algorithm taking account of the varying quantities for elements deployed at the STAR-RIS and the BS, as well as different maximum transmitting powers. From this figure, it is observed that the objective function values monotonically increase with the number of iterations. Additionally, the proposed algorithm consistently exhibits rapid convergence to a stable value in fewer than $10$ iterations across different scenarios, suggesting that the convergence of the algorithm can be reliably ensured.

\begin{figure}[ht]
	\centering
	\includegraphics[scale=0.45]{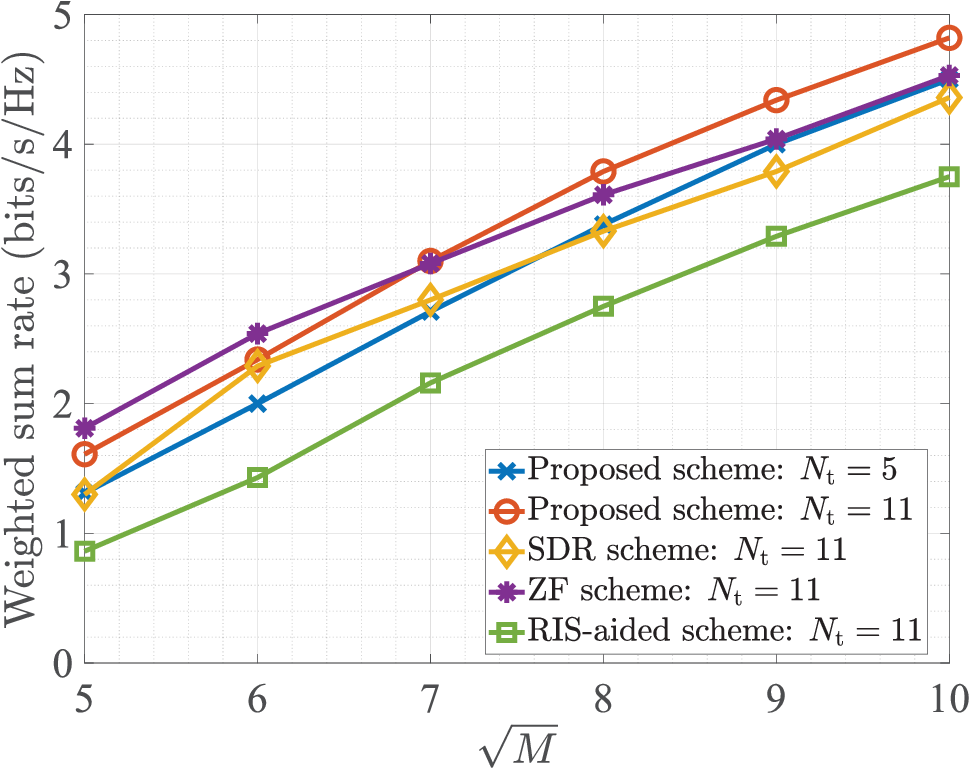}\\
	\caption{Weighted sum rate versus the number of elements equipped at the STAR-RIS with $P_1=P_0=0.5$, $\omega_1=\omega_2=0.5$, $P_\mathrm{tmax}=30$ dBm,  $\lambda=2$  and different $N_\mathrm{t}$.}\label{fig:WS_vs_M}
\end{figure}
The influence of the number of elements deployed at the STAR-RIS on the weighted sum rate is explored in Fig. \ref{fig:WS_vs_M} considering varying number of antennas. In particular, it is observed that all cases exhibit a rising pattern with the growth of $M$, while the rate of increase has undergone a gradual decline. This is due to the fact that more elements can provide more degrees of freedom to reconfigure the wireless propagation environment.
Moreover, an interesting phenomenon can be found that the ZF scheme demonstrates superior performance gains when the value of $M$ is low, whereas the proposed scheme surpasses the ZF scheme with identical configurations as the number of elements increases. This is because the ZF scheme is fundamentally designed with the objective of eliminating interference among users. Therefore, even when the control capacity of STAR-RIS is limited, it can still achieve significant performance gains. In contrast, the proposed scheme faces challenges in eliminating interference among users when the control capacity of STAR-RIS is limited, i.e., when $M$ is small. With an increase in $M$, the adjustment capacity of STAR-RIS over incoming signals is enhanced, enabling not only effective interference elimination among users but also improving the quality of received signals for different users, leading to further enhancement in communication rate. Additionally, the proposed scheme shows the significant advantage by comparing the simulation results with the SDR scheme under same conditions, which demonstrates the efficiency of the proposed algorithm. Note that the traditional RIS-assisted system shows minimal performance improvement when compared to other baseline schemes, which is even worse than the proposed scheme with more stringent system conditions (i.e., $N_\mathrm{t}=5$). This shows the notable capability of the STAR-RIS in effectively implementing full-space modulation to signals.

Next, we investigate the varying trend of the weighted sum rate with the number of antennas at the BS by considering the $P_1=P_0=0.5$, $\omega_1=\omega_2=0.5$, $\lambda=2$ and different number of elements equipped at the STAR-RIS. Specifically,  a similar pattern of performance can still be observed, where the weighted sum rate increases as the growth of $N_{\mathrm{t}}$ across all scenarios, with a decreasing growth rate. It is worth noting that the suggested scheme exhibits the highest performance gain for the communication systems when compared to three other baseline schemes operating under identical conditions, thereby demonstrating the efficacy of both the proposed STAR-RIS-assisted scheme and designed algorithm.
\vspace{-2mm}
\begin{figure}[ht]
	\centering
	\includegraphics[scale=0.45]{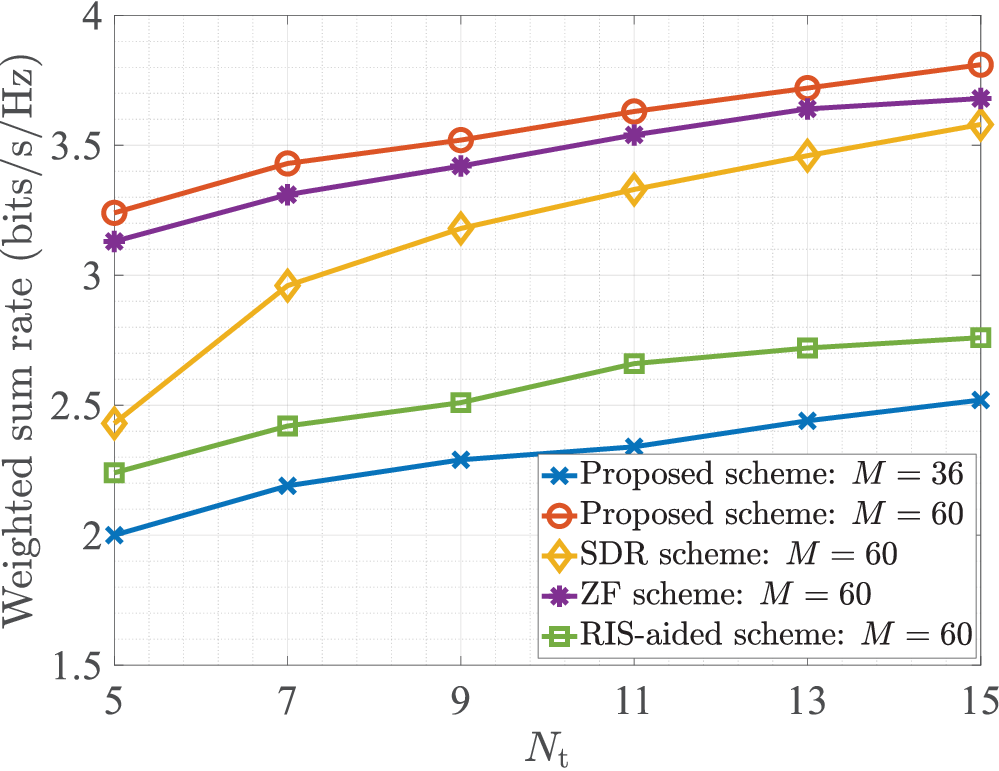}\\
	\caption{Weighted sum rate versus the number of antennas with $P_1=P_0=0.5$, $\omega_1=\omega_2=0.5$, $P_\mathrm{tmax}=30$ dBm,  $\lambda=2$  and different $M$.}\label{fig:WS_vs_Nt}
\end{figure}

\begin{figure}[ht]
	\centering
	\includegraphics[scale=0.45]{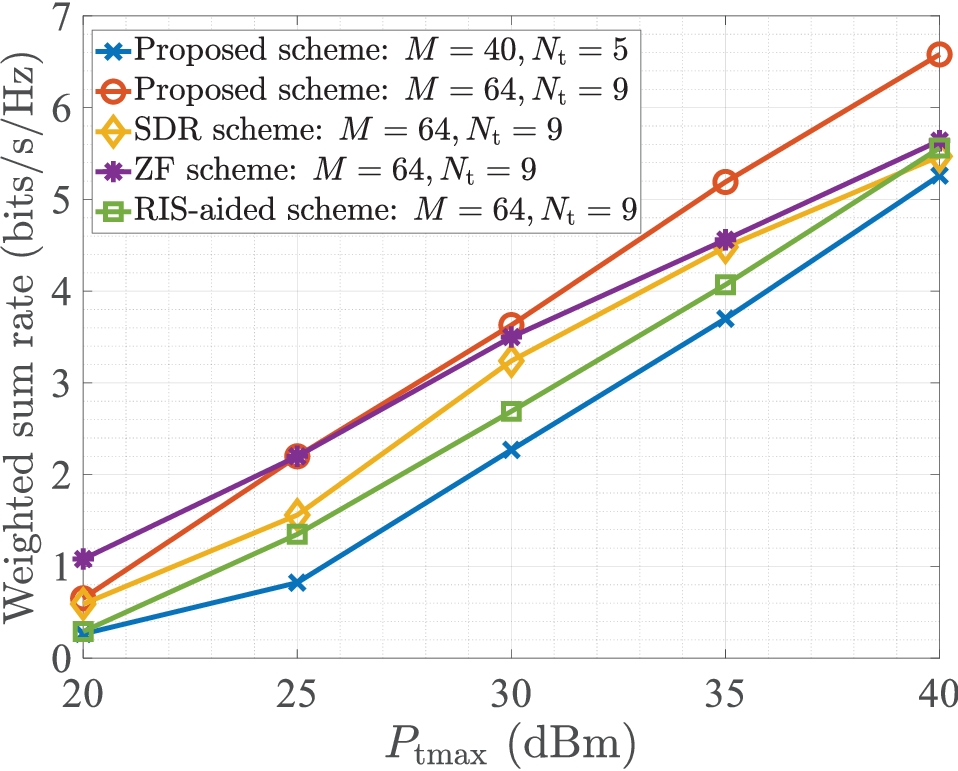}\\
	\caption{Weighted sum rate versus the maximum transmitting power with $P_1=P_0=0.5$, $\omega_1=\omega_2=0.5$,  $\lambda=2$, and different $N_\mathrm{t}$ and $M$.}\label{fig:WS_vs_Ptmax}
\end{figure}

The changed correlation of the weighted sum rate as the maximum transmitting power is explored in Fig. \ref{fig:WS_vs_Ptmax} with different $M$ and $N_\mathrm{t}$.  Specifically, two cases are selected to implement the proposed scheme, i.e., $( M=40, N_\mathrm{t}=5)$ and $( M=64, N_\mathrm{t}=9)$, in order to authenticate the general applicability of the proposed approach. For these benchmark schemes, $( M=64, N_\mathrm{t}=9)$ is utilized to implement them. We can find that the weighted sum rate experiences a swift growth as the power budget increases, with the enhanced rate showing a consistent upward trajectory. Generally, the presented scheme demonstrates notable improvement in performance for communication systems in comparison to other  baseline schemes. And the scheme facilitated by the conventional RIS shows the least favourable performance outcomes when subjected to the same conditions.

\begin{figure}[ht]
	\centering
	\includegraphics[scale=0.45]{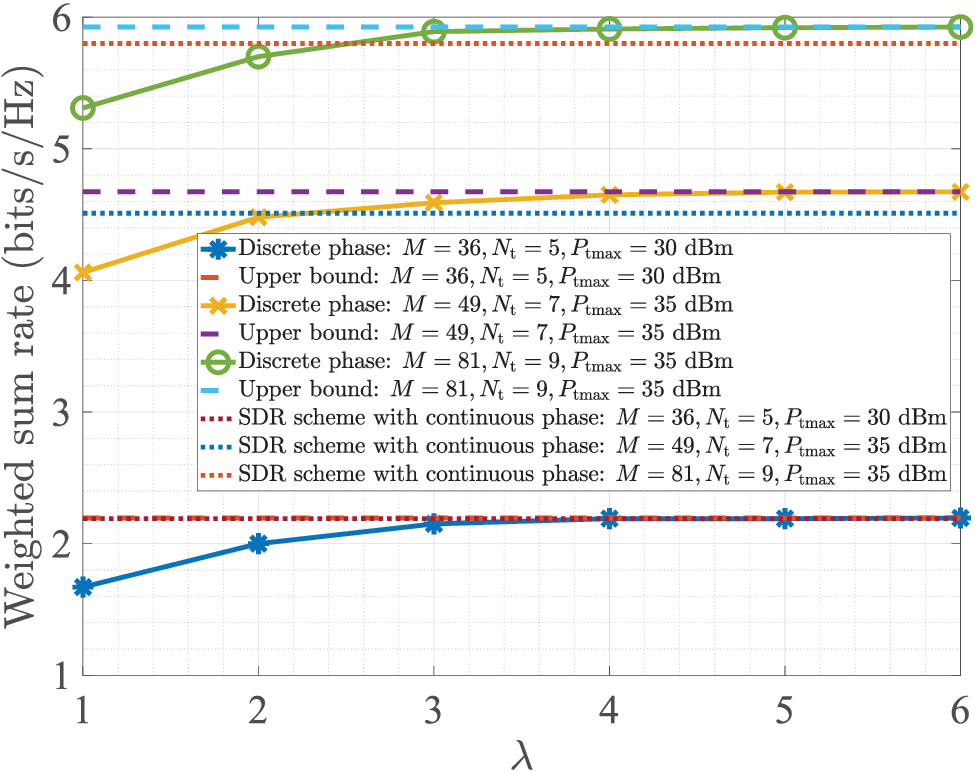}\\
	\caption{Weighted sum rate versus the number of quantification bits with $P_1=P_0=0.5$, $\omega_1=\omega_2=0.5$, and different $N_\mathrm{t}$, $M$ and $P_\mathrm{tmax}$.}\label{fig:WS_vs_lambda}
\end{figure}

In Fig. \ref{fig:WS_vs_lambda}, we plots the weighted sum rate versus the number of quantification bits taking into account  different $N_\mathrm{t}$, $M$ and $P_\mathrm{tamx}$. It is worth noting that the weighted sum rate gradually increases and converge to a specific point as the growth of $\lambda$ for all considering scenarios. This is due to the fact that the enhanced capacity of the STAR-RIS to provide a greater number of phase shifts for signal modulation contributes to the growth of performance gain. Furthermore, it is also observed that a minimum $4$-bit qualification is necessary to approach the performance limit for all scenarios, which is may because the STAR-RIS may require more precise phase-shifts to enhance signal quality for authorized users and simultaneously minimize signal leakage in full-space service convergence. Additionally, we also present the simulation results achieved by the SDR scheme with continuous phase-shifts. According to these results, it is observed that the suggested scheme utilizing discrete phase-shifts demonstrates a slight performance advantage over the SDR scheme employing continuous phase-shifts, particularly as the value of $M$ escalates. This indicates that the proposed scheme demonstrates substantial potential for application in practical STAR-RIS implementation.
\begin{figure}[ht]
	\centering
	\includegraphics[scale=0.40]{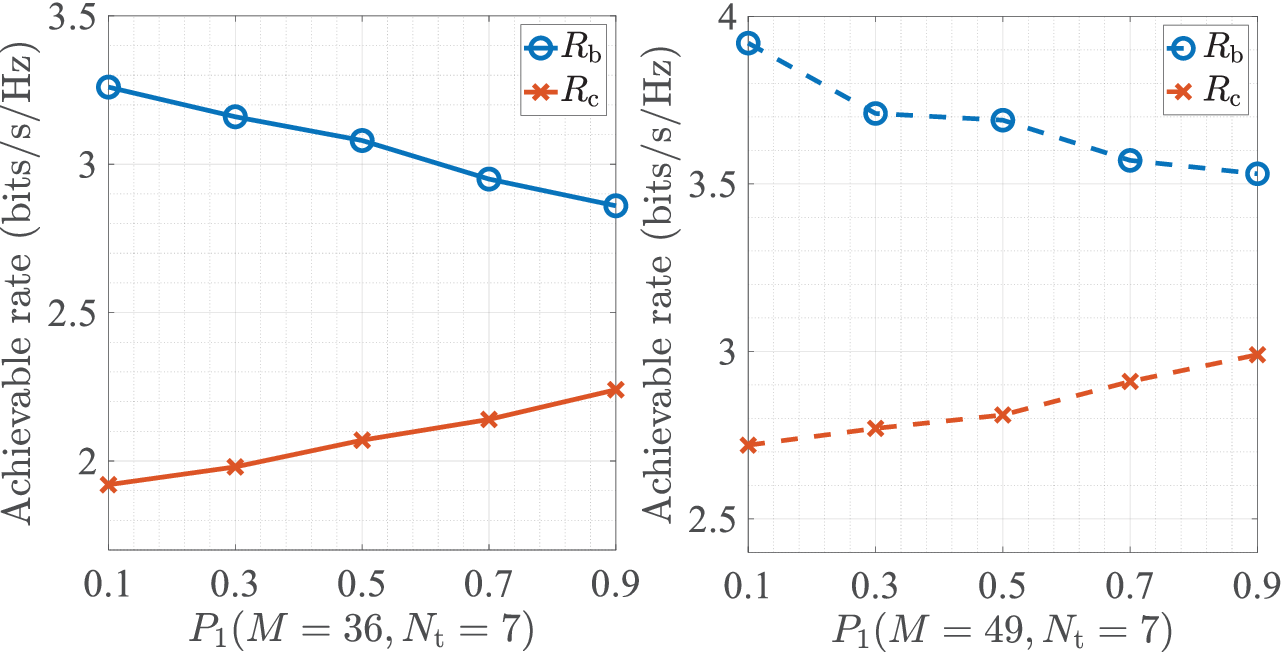}\\
	\caption{Achievable rate versus $P_1$with  $\omega_1=\omega_2=0.5$, $P_\mathrm{tmax}=30$ dBm,  $\lambda=2$  and different $N_\mathrm{t}$ and $M$.}\label{fig:P1_vs_rate}
\end{figure}

The achievable rates at Bob and Carol versus the probability of Eve being situated in reflection region of the STAR-RIS is investigated, as shown in Fig. \ref{fig:P1_vs_rate}. Specifically, two cases are considered to explore this correlation. From this figure, we observe an upward trend in the achievable rate at Bob as the value of $P_1$ increases, whereas Carol's achievable rate shows a contrasting trend by decreasing. This reason can be summarized as: As the probability of Eve appearing in the reflection zone of STAR-RIS increases with the increase of $P_1$, STAR-RIS will allocate more reflection beamforming resources to suppress information leakage in the reflection zone, in order to ensure the security of Bob's information. Hence, Bob's attainable rate shows a downward trend. Conversely,  the passive beamforming resources allocated for minimizing the information leakage will decrease as $P_1$ increases in transmission region. Consequently, Carol's achievable rate demonstrates an upward trend.
\begin{figure}[ht]
	\centering
	\includegraphics[scale=0.63]{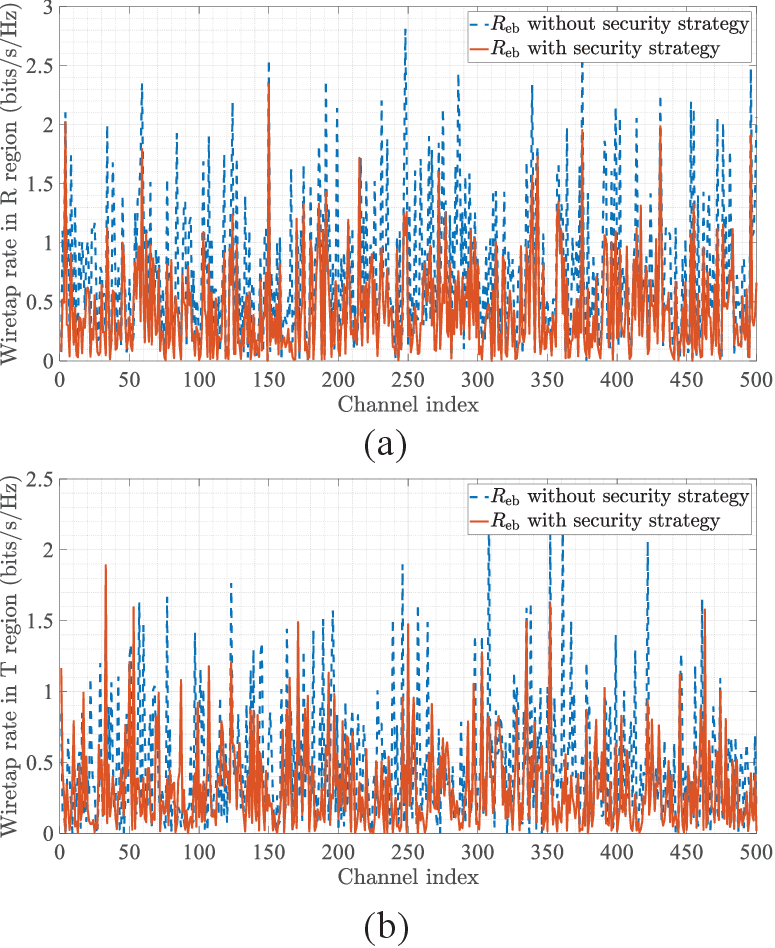}\\
	\caption{The effectiveness of the proposed scheme in suppression eavesdropping rate with $P_1=P_0=0.5$, $\omega_1=\omega_2=0.5$, $N_\mathrm{t}=9$, $M=64$ and $P_\mathrm{tmax}=30$ dBm: (a) Reflection region; (b) Transmission region.}\label{fig:R_eb_vs_channel}
\end{figure}

To validate the effectiveness of the proposed PLS scheme in suppressing information leakage across the full-space coverage region of STAR-RIS-supported wireless networks, we first generate $500$ random eavesdropper's channels and compute the corresponding wiretapping rates with and without the security strategy. This is done based on optimized active and passive beamforming solutions. In particular, Fig.
 \ref{fig:R_eb_vs_channel} (a) and (b) present the wiretap rates in R region and T region, respectively.
The results from the graph show that the eavesdropping rates considering different Eve's CSI in R and T regions are significantly suppressed with the assistance of the proposed STAR-RIS-assisted PLS scheme when compared to the results obtained without considering security strategy, which demonstrates that the proposed robust full-space PLS scheme can effectively restrain Eve's eavesdropping for communications between the BS and Bob. Furthermore, we can find that these wiretap rates remain notably elevated within certain Eve's channels, primarily attributed to the high level of coupling exhibited in these channels with Bob's communication channel. These findings from the simulation results also indicate that the reflection area is subject to a higher vulnerability to eavesdropping when compared to the transmission region, as evidenced by the wiretap rate values.

\section{Conclusion and Prospect}\label{sec:S6}
This paper proposes a robust full-space PLS transmission scheme to protect the information security of the STAR-RIS-assisted communication systems. Specifically, we first analytically derive the asymptotic form of the average security rate using the large system analysis techniques considering the challenging scenario where only the eavesdropper's statistical CSI is available at the BS and there is ambiguity regarding the eavesdropper's location in either R region or T region of the STAR-RIS. Then, an optimization problem is outlined with the objective of enhancing the combined weighted sum rate of the mean security rate at Bob and the attainable rate of Carol through the concurrent optimization of active and passive beamforming variables. This is subject to constraints on transmit power, equality amplitude constraints, and discrete phase-shift constraints. A novel iterative algorithm based on the MMSE method and CEO approach is proposed to effectively address this non-convex problem. The effectiveness of the proposed PLS scheme and algorithm has been validated through a comprehensive comparative analysis with a large number of simulation results against three baseline schemes including two benchmark algorithms and the conventional RIS-assisted scheme. \textcolor{blue}{
%Note that in designing the PLS scheme and corresponding algorithms, our work 
In this paper, we adopt a practical approach that considers both the eavesdropping risks faced by STAR-RIS-assisted communications and the hardware constraints encountered in practical deployments of STAR-RIS, %. Consequently, the proposed scheme not only effectively enhances the system's confidentiality performance but also accounts for hardware feasibility, providing
providing a viable theoretical framework for the practical application of STAR-RIS in wireless communication systems.}

\textcolor{blue}{This study focuses on the ES protocol of STAR-RIS and considers an eavesdropping scenario with a single eavesdropper. In future works, it would be valuable to explore alternative protocols of STAR-RIS, such as mode selection and time switching, as well as to extend the scenarios with multiple eavesdroppers. Additionally, integrating active or hybrid STAR-RIS configurations into the proposed PLS scheme %, and considering the deployment of unmanned aerial vehicles (UAVs) 
to further enhance secure communications, %represent important avenues for 
deserve further investigations. These extensions are expected to broaden the applicability and robustness of STAR-RIS-assisted systems in practical and dynamic environments.}

\appendices
\section{Proof of Theorem 1}\label{app_1}
The first-order partial derivatives of the Lagrange function in \eqref{eq_active_trans_La} w.r.t $\mathbf{w}_\mathrm{b}$ and $\mathbf{w}_\mathrm{b}$ can be derived as

\begin{align}
&\frac{\partial\mathcal{L}}{\partial{\mathbf{w}_\mathrm{b}}}=2\varrho\mathbf{w}_\mathrm{b}+\frac{2}{\ln2}\omega_2W_\mathrm{c}^{(t-1)}\left|u_\mathrm{c}^{(t-1)}\right|^2\mathbf{A}\mathbf{w}_\mathrm{b}+2\omega_1\times\notag\\
	&~\Bigg(\bigg(\frac{W_1^{(t-1)}\left|u_1^{(t-1)}\right|^2}{\ln2}\mathbf{B}\mathbf{w}_\mathrm{b}-\frac{W_1^{(t-1)}}{\ln2}\mathbf{H}_\mathrm{BR}^H\mathbf{U}_\mathrm{r}^H\mathbf{h}_\mathrm{rb}u_1^{(t-1)}\bigg)\notag\\
	&~+\frac{l_\mathrm{re}P_1W_4^{(t-1)}}{\sigma_\mathrm{e}^2\ln2}\mathbf{C}_\mathrm{r}\mathbf{w}_\mathrm{b}+\frac{l_\mathrm{re}P_0W_5^{(t-1)}}{\sigma_\mathrm{e}^2\ln2}\mathbf{C}_\mathrm{t}\mathbf{w}_\mathrm{b}\Bigg).\\	
	&\frac{\partial\mathcal{L}}{\partial{\mathbf{w}_\mathrm{c}}}=2\varrho\mathbf{w}_\mathrm{c}+\frac{2\omega_2W_\mathrm{c}^{(t-1)}}{\ln2}\Big(\left|u_\mathrm{c}^{(t-1)}\right|^2\mathbf{A}\mathbf{w}_\mathrm{c}\notag\\
	&~-\mathbf{H}_\mathrm{BR}^H\mathbf{U}_\mathrm{t}^H\mathbf{h}_\mathrm{rc}u_\mathrm{c}^{(t-1)}\Big)+2\omega_1\Bigg(\frac{W_1^{(t-1)}\left|u_1^{(t-1)}\right|^2}{\ln2}\mathbf{B}\mathbf{w}_\mathrm{c}\notag\\
	&~+\frac{P_1W_2^{(t-1)}}{\ln2}\left(l_\mathrm{re}\mathbf{D}\mathbf{w}_\mathrm{c}-\sqrt{l_\mathrm{re}}\mathbf{H}_\mathrm{BR}^H\mathbf{U}_\mathrm{r}^H\mathbf{u}_2^{(t-1)}\right)\notag\\
	&~+\frac{P_0W_3^{(t-1)}}{\ln2}\left(l_\mathrm{re}\tilde{\mathbf{D}}\mathbf{w}_\mathrm{c}-\sqrt{l_\mathrm{re}}\mathbf{H}_\mathrm{BR}^H\mathbf{U}_\mathrm{t}^H\mathbf{u}_3^{(t-1)}\right)\notag\\
	&~+\frac{l_\mathrm{re}P_1W_4^{(t-1)}}{\sigma_\mathrm{e}^2\ln2}\mathbf{C}_\mathrm{r}\mathbf{w}_\mathrm{c}+\frac{l_\mathrm{re}P_0W_5^{(t-1)}}{\sigma_\mathrm{e}^2\ln2}\mathbf{C}_\mathrm{t}\mathbf{w}_\mathrm{c}\Bigg).
\end{align}
where
\begin{itemize}
	\item$\mathbf{A}=\mathbf{H}_\mathrm{BR}^H\mathbf{U}_\mathrm{t}^H\mathbf{h}_\mathrm{rc}\mathbf{h}_\mathrm{rc}^H\mathbf{U}_\mathrm{t}\mathbf{H}_\mathrm{BR},~ \mathbf{C}_\mathrm{r}=\mathbf{H}_\mathrm{BR}^H\mathbf{U}_\mathrm{r}^H\mathbf{U}_\mathrm{r}\mathbf{H}_\mathrm{BR}$,
			\vspace{2mm}
\item $\mathbf{B}=\mathbf{H}_\mathrm{BR}^H\mathbf{U}_\mathrm{r}^H\mathbf{h}_\mathrm{rb}\mathbf{h}_\mathrm{rb}^H\mathbf{U}_\mathrm{r}\mathbf{H}_\mathrm{BR},
	\vspace{1mm}~ \mathbf{C}_\mathrm{t}=\mathbf{H}_\mathrm{BR}^H\mathbf{U}_\mathrm{r}^H\mathbf{U}_\mathrm{t}\mathbf{H}_\mathrm{BR},$
			\vspace{2mm}
		\item$\mathbf{D}=\mathbf{H}_\mathrm{BR}^H\mathbf{U}_\mathrm{r}^H\mathbf{u}_2^{(t-1)}\left(\mathbf{u}_2^{(t-1)}\right)^H\mathbf{U}_\mathrm{r}\mathbf{H}_\mathrm{BR},$ \item$\tilde{\mathbf{D}}=\mathbf{H}_\mathrm{BR}^H\mathbf{U}_\mathrm{t}^H\mathbf{u}_3^{(t-1)}\left(\mathbf{u}_3^{(t-1)}\right)^H\mathbf{U}_\mathrm{t}\mathbf{H}_\mathrm{BR}.$
			\vspace{2mm}
\end{itemize}
Let $\frac{\partial\mathcal{L}}{\partial{\mathbf{w}_\mathrm{b}}}=0$ and $\frac{\partial\mathcal{L}}{\partial{\mathbf{w}_\mathrm{c}}}=0$, we have
\begin{align}
	&	\mathbf{w}_\mathrm{b}^\mathrm{opt}(\varrho)=\mathbf{G}_1^{-1}\mathbf{G}_2, ~\mathbf{w}_\mathrm{c}^\mathrm{opt}(\varrho)=\widehat{\mathbf{G}}_1^{-1}\widehat{\mathbf{G}}_2,
\end{align}
where
\begin{itemize}
\item$\mathbf{G}_1=\varrho\mathbf{I}_{N_\mathrm{t}\times N_\mathrm{t}}+\frac{\omega_2W_\mathrm{c}^{(t-1)}\left|u_\mathrm{c}^{(t-1)}\right|^2}{\ln2}\mathbf{A}+\mathbf{B}\times\\
    \frac{\omega_1W_1^{(t-1)}\left|u_1^{(t-1)}\right|^2}{\ln2}+\frac{\omega_1l_\mathrm{re}P_1W_4^{(t-1)}}{\sigma_\mathrm{e}^2\ln2}\mathbf{C}_\mathrm{r}+\frac{\omega_1l_\mathrm{re}P_0W_5^{(t-1)}}{\sigma_\mathrm{e}^2\ln2}\mathbf{C}_\mathrm{t},$
		\vspace{2mm}
\item$\mathbf{G}_2=\frac{\omega_1W_1^{(t-1)}}{\ln2}\mathbf{H}_\mathrm{BR}^H\mathbf{U}_\mathrm{r}^H\mathbf{h}_\mathrm{rb}u_1^{(t-1)},$
			\vspace{2mm}
\item$\widehat{\mathbf{G}}_1=\varrho\mathbf{I}_{N_\mathrm{t}\times N_\mathrm{t}}+\frac{\omega_2W_\mathrm{c}^{(t-1)}}{\ln2}\left|u_\mathrm{c}^{(t-1)}\right|^2\mathbf{A}+
    \frac{\omega_1W_1^{(t-1)}}{\ln2}\times\\\left|u_1^{(t-1)}\right|^2\mathbf{B}+\frac{\omega_1P_1W_2^{(t-1)}l_\mathrm{re}}{\ln2}\mathbf{D}+
    \frac{\omega_1P_0W_3^{(t-1)}l_\mathrm{re}}{\ln2}\tilde{\mathbf{D}}+\\\frac{\omega_1l_\mathrm{re}P_1W_4^{(t-1)}}{\sigma_\mathrm{e}^2\ln2}\mathbf{C}_\mathrm{r}+\frac{\omega_1l_\mathrm{re}P_0W_5^{(t-1)}}{\sigma_\mathrm{e}^2\ln2}\mathbf{C}_\mathrm{t},$
	\vspace{2mm}		\item$\widehat{\mathbf{G}}_2=\frac{\omega_2W_\mathrm{c}^{(t-1)}}{\ln2}\mathbf{H}_\mathrm{BR}^H\mathbf{U}_\mathrm{t}^H\mathbf{h}_\mathrm{rc}u_\mathrm{c}^{(t-1)}+\frac{\omega_1P_1W_2^{(t-1)}\sqrt{l_\mathrm{re}}}{\ln2}\\\times\mathbf{H}_\mathrm{BR}^H\mathbf{U}_\mathrm{r}^H\mathbf{u}_2^{(t-1)}+\frac{\omega_1P_0W_3^{(t-1)}\sqrt{l_\mathrm{re}}}{\ln2}\mathbf{H}_\mathrm{BR}^H\mathbf{U}_\mathrm{t}^H\mathbf{u}_3^{(t-1)}.$
			\vspace{2mm}
\end{itemize}
\section{Proof of Theorem 2}\label{app_2}
The Lagrange function of optimization problem \eqref{eq_entropy} can be expressed as
\begin{align}
	\mathcal{L}(\boldsymbol{\Upsilon},\xi)=-\frac{1}{K}\sum_{k=1}^{K^\mathrm{elite}}\ln	\mathscr{F}(\boldsymbol{\Xi}_{[k]};\boldsymbol{\Upsilon})+\sum_{p=1}^{2M}\xi_p(\sum_{q=1}^{Q}P_{pq}-1).
\end{align}
The first-order partial derivatives of the Lagrange function w.r.t $P_{pq}$, $\sigma_{m}$ and $\mu_{m}$ can be derived as
\begin{align}
	&\frac{\partial\mathcal{L}}{\partial P_{pq}}=\xi_p-\frac{1}{K}\sum_{k=1}^{K^\mathrm{elite}}\frac{\mathbbm{1}_{\{\mathcal{I}_{p}(\boldsymbol{\phi}_{[k]})=\varphi_{q}\}}}{P_{pq}},\\
	&\frac{\partial\mathcal{L}}{\partial\sigma_{m}}=-\frac{1}{K}\sum_{k=1}^{K^\mathrm{elite}}\frac{(\mathcal{I}_m(\boldsymbol{\beta}^\mathrm{r}_{[k]})-\mu_m)^2-K^\mathrm{elite}\sigma_{m}^2}{\sigma_{m}^3},\\
	&\frac{\partial\mathcal{L}}{\partial\mu_{m}}=-\frac{1}{K}\sum_{k=1}^{K^\mathrm{elite}}\frac{\mathcal{I}_m(\boldsymbol{\beta}^\mathrm{r}_{[k]})-K^\mathrm{elite}\mu_m}{\sigma_{m}^2}.
\end{align}
Let $\frac{\partial\mathcal{L}}{\partial P_{pq}}=0$, $\frac{\partial\mathcal{L}}{\partial\sigma_{m}}=0$ and $\frac{\partial\mathcal{L}}{\partial\mu_{m}}=0$, we have
\begin{align}
	&P_{pq}=\frac{\sum_{k=1}^{K^\mathrm{elite}}\mathbbm{1}_{\{\mathcal{I}_p(\boldsymbol{\phi}_{[k]})=\varphi_{q}\}}}{K\xi_p}, ~\forall p\in\mathcal{P}, ~q\in\mathcal{Q},\\
	&\xi_p=\frac{K^\mathrm{elite}}{K}, ~\forall p\in\mathcal{P},
\end{align}	
\begin{align}	
	&\sigma_{m}=\sqrt{\frac{\sum_{k=1}^{K^\mathrm{elite}}\Big(\mathcal{I}_m(\boldsymbol{\beta}_{[k]}^\mathrm{r})-\mu_m\Big)^2}{K^\mathrm{elite}}},  ~\forall m\in\mathcal{M},\\
	&\mu_{m}=\frac{\sum_{k=1}^{K^\mathrm{elite}}\mathcal{I}_m(\boldsymbol{\beta}_{[k]}^\mathrm{r})}{K^\mathrm{elite}},  ~\forall m\in\mathcal{M}.
\end{align}
Substituting the expression of $\xi_p$ into the expression of $P_{pq}$, we can easily obtain the results presented in \eqref{eq_P_update}- \eqref{eq_mu_update}.

\ifCLASSOPTIONcaptionsoff
\newpage
\fi
\bibliographystyle{IEEEtran}
\bibliography{PLS}

\end{document}